\documentclass[preprint,11pt]{elsarticle}
\usepackage{setspace} 
\usepackage{graphicx}
\usepackage{amssymb}
\usepackage[ruled,vlined]{algorithm2e}

\newtheorem{defi}{Definition}
\newtheorem{lemma}{Lemma}
\newtheorem{theorem}{{\bf Theorem}}
\newtheorem{obs}{{\bf Observation}}

\newtheorem{corollary}{{\bf Corollary}}

\begin{document}
\begin{frontmatter}
\title{Leader Election and Gathering for Asynchronous Transparent Fat Robots without Chirality}

\author{Sruti Gan Chaudhuri and Krishnendu Mukhopadhyaya}
\address{ACM Unit, Indian Statistical Institute,\\203 B.T. Road, Kolkata - 700108, India.
\\Email: sruti\_r@isical.ac.in, krishnendu@isical.ac.in}

\vspace{-50pt}
\begin{abstract}

\small
\emph{Gathering} 
of autonomous mobile robots is a well known and challenging research problem for a system of multiple mobile robots. Most
of the existing works consider the robots to be dimensionless or points.
Algorithm for gathering \emph{fat robots} (a robot represented as a unit
disc) has been reported for {\it three} and {\it four} robots. This paper proposes a distributed algorithm which deterministically gathers $n$
$(n \ge 5)$ asynchronous, fat robots. The
robots are assumed to be transparent and they have full visibility. Since the robots are unit discs, they can not overlap; we define a {\it gathering pattern} for them.
The robots are initially considered to be stationary. A robot is visible in its motion. The robots do not 
store past actions.
They are anonymous and can not be distinguished by their appearances and do not have common coordinate system or chirality. The robots do not communicate 
through message passing.  

In the proposed {\it gathering} algorithm one robot moves at a time towards its destination. The robot which moves, is selected in such
a way that, it will be the only robot eligible to move, until it reaches its
destination. In case of a tie, this paper proposes a {\it leader election} algorithm which produces an ordering 
of the robots and the first robot in the ordering becomes the leader. The ordering is unique in the sense that, each robot, characterized by its location, agrees on the same ordering. 
We show that if a set of robots can be ordered then they
can gather deterministically. The paper also characterizes the cases, where ordering is not
possible. 

This paper also presents an important fact that, if leader election is possible then {\it gathering pattern} formation is possible even with no chirality. 
It is known that, pattern formation and leader election are equivalent, for $n \ge 4$ robots when the robots agree on chirality. This paper proposes an algorithm for formation of the {\it gathering pattern} 
for transparent fat robots without chirality. 

\begin{keyword}
 Asynchronous, fat robots, gathering, leader election, chirality. 
\end{keyword}
\end{abstract}

\end{frontmatter}

\section{Introduction}
\label{introduction}

A \emph{Robot Swarm} \cite{peleg2005,peleg2009} is a system of multiple autonomous
mobile robots engaged in some collective task. In hostile environments, it may
be desirable to employ large groups of low cost robots to
perform various tasks cooperatively. This approach is more resilient to malfunction and
more configurable than a single high cost robot. \emph{Swarm robotics},
pioneered by C. W. Reynolds \cite{reynolds1987}, is a novel approach to coordinate a
large number of robots. The idea is inspired by the observation of social
insects. They are known to coordinate their actions to execute a task that is
beyond the capability of a unit.

The field of swarm robotics has been enriched by many researchers adopting
different approaches for swarm aggregation, navigation, coordination and
control. Mobile robots can move in the physical world and interact with each other.
Geometric problems are inherent to such multiple cooperative mobile robot
systems and have been well studied \cite{acm2004, IEEEJOR1999, celi2003, celi2002, 
cohen2004, cohen2005, disc2010, flocchini2001, gordon2008, peleg2005, prencipe2007}. 
Multiple robot path planning, moving to (and maintaining) formation and pattern generation are some important geometric
problems in swarm robotics. Multiple robot path planning may deal with problems
like finding non-intersecting paths for mobile robots \cite{Fuj91}. The
formation and marching problems require multiple robots to form up and move in a
specified pattern. These problems are also interesting in terms of
distributed algorithms. Formation and marching problems may act as useful
primitives for larger tasks, like, moving a large object by a group of robots
\cite{SB93} or distributed sensing \cite{WB88}. Pattern generation in Cellular
Robotic Systems (CRS) \cite{ben1988} is related to pattern formation problem by
mobile robots.

This paper addresses a very well known and challenging problem involving robot
swarms, namely \emph{Gathering}. The objective is to collect multiple autonomous mobile
robots into a point or a small region. The choice of the point is not fixed in
advance. Initially the robots are stationary and in arbitrary positions. Gathering
problem is also referred to as {\it Point Formation}, {\it Convergence},
\emph{Homing} or {\it Rendezvous} \cite{prencipe2006}.

\section{Earlier Works}
\label{earlier_works}
A pragmatic view of swarm robots asks for  distributed environment. Several interesting works 
have been carried out by researchers \cite{acm2004,cohen2004,peleg2009,flocchini2000, 
gordon2008,katayama2007,klasing2006} on distributed algorithms for mobile robots. 
A simple basic model called \emph{weak model} \cite{peleg2007,prencipe2006} is 
popular in the literature. The world of the robots consists of the 
infinite plane and multiple robots living on it. The robots were considered dimensionless 
or points. All robots are autonomous, homogeneous and perform the same algorithm ({\it
Look-Compute-Move} cycle) \cite{peleg2007}. In \emph{look} state, a robot takes a snapshot of its
surroundings, within its range of vision. It then
executes an algorithm for computing the destination in \emph{compute} state. The
algorithm is same for all robots. In \emph{move} state, the robot moves to the
computed destination. The robots are oblivious (memoryless). They do not
preserve any data computed in the previous cycles. There is no explicit
communication between the robots. The robots coordinate by means of observing
the positions of the other robots on the plane. A robot
is always able to see another robot within its visibility radius or range (may
be infinite). Two different models have been used for robots' movement \cite{prencipe2001}. Under
the SYM model, the movement of a robot was considered to be instantaneous,
i.e., when a robot is moving, other robots can not see it. Later, that model has
been modified to CORDA model where the movement of a robot is
not instantaneous. A robot in motion is visible. The CORDA model is a
better representation of the real world. The robots may or may not be
synchronized. Synchronous robots execute their cycles together. In such a
system, all robots get the same view. As a result, they compute on same data. In
the more practical asynchronous model, there is no such guarantee. By the time a
robot completes its computation, several of the robots may have moved from the
positions based on which the computation is made. The robots may have a common
coordinate system or individual coordinate systems having no common orientation
or scale \cite{peleg2007}.

The problem of gathering multiple robots has been studied on the basic model of
robotic system with different additional assumptions. Prencipe \cite{prencipe2007}
observed that gathering is not always possible for asynchronous robots. However,
instead of meeting at a single point, if the robots are asked to move very
close to each other, then the problem can be solved by computing the center of
gravity of all robots and moving the robots towards it. Under the weak model,
robots do not have any memory. As they work asynchronously and act
independently, once a robot moves towards the center of gravity, the center of
gravity also changes. Moving towards center of gravity does not gather robots to 
a single point. However, this method makes sure that the
robots can be brought as close as one may wish. 

A possible solution to this problem is to choose a point which, unlike center of gravity,
is invariant with respect to robots' movement. One such point in the plane is
the point which minimizes the sum of distance between itself and all the robots.
This point is called the \emph{Weber or Fermat or Torricelli} point. It does not
change during the robots' movement, if the robots move only towards this point. 
However, the \emph{Weber} point is not
computable for higher number of robots (greater than 4) \cite{bajaj1988}. Thus,
this approach also can not be used to solve the gathering problem. If common
coordinate system is assumed, instead of individual coordinate systems, then the
Gathering problem is solvable
even with limited visibility \cite{flocchini2001}. If the robots are synchronous
and their movements are instantaneous, then the gathering problem is solvable
even with limited visibility \cite{IEEEJOR1999}. Cieliebak et al. \cite{celi2003}, show that
there are solutions for gathering without synchronicity if the visibility is
unlimited. One approach is by {\it multiplicity} detection. If
multiple robots are at the same point then the point is said to have {\it
strict multiplicity}. However, the problem is not solvable for {\it two} robots
even with {\it multiplicity} detection \cite{prencipe2007}. There are some
solutions for {\it three} and {\it four} robots. For more than {\it four}
robots, there are two algorithms with restricted sets of initial configurations
\cite{celi2002}. In the first algorithm, the robots are initially in a {\it
bi-angular configuration}. In such a configuration, there exists a point $c$ and two angles $a$ and $b$ such that the angle between
any two adjacent robots is either $a$ or $b$ and the angles alternate. The
second algorithm works if the initial configuration of the robots do not form
any regular $n$-gon. Prencipe \cite{prencipe2007} reported that there exists no
deterministic oblivious algorithm that solves the gathering problem in a finite
number of cycles for a set of $n\geq2$ robots. {\it Convergence} of multiple
robots is studied by Peleg and Cohen \cite{cohen2005} and they proposed a
gravitational algorithm in fully asynchronous model for convergence of any
number of robots.

A dimensionless robot is unrealistic.
Czyzowicz et al.,\cite{czy2009} extend the weak model by replacing the point robots
by unit disc robots. They called these robots as \emph{fat robots}. The methods
of gathering of three and four fat robots have been described by them. The paper considers partial
visibility and presents several procedures to avoid different situations which
cause obstacles to gathering. We proposed a deterministic gathering algorithm for $n$ $(n \ge 5)$
fat robots \cite{km2010}. The robots are assumed to be transparent in order to achieve
full visibility.

Having {\it fat robots}, we first define the {\it gathering pattern} to be formed by the robots. Therefore, 
{\it gathering pattern} formation becomes a special case of {\it pattern formation} of mobile robots which is also a challenging problem for mobile robots. Recent works \cite{disc2010, dieu2010} 
shows that pattern formation and leader election are equivalent for $n \ge 4$ robots. 
However, this work \cite{disc2010} considered the robots to have common handedness or chirality. We propose a gathering 
algorithm which assumes no chirality and use the leader election technique in order to form gathering pattern. We also show that if leader 
election is possible then formation of the {\it gathering pattern} is possible even with no chirality. 
Section \ref{model} describes the model used in this paper and
presents an overview of the problem. Then we move to the solution
approach. Section \ref{char} characterizes the geometric configurations of the robots for gathering. 
Section \ref{algo} presents the {\it leader election} and {\it gathering} algorithms. Finally section \ref{con} summarizes the 
contributions of the paper and concludes.

\section{Robot Model and Overview of the Problem}
\label{model}
We use the basic structure of {\it weak model} \cite{peleg2007,prencipe2006} and add
some extra features which extend the model. Let $R= \{r_1, r_2, \ldots, r_n\}$ be a set of fat 
robots. A robot
is represented by its center, i.e., by $r_i$ we mean a robot whose center is
$r_i$. The following assumptions describe the system of robots deployed on the 2D
plane:
\begin{itemize}
\item Robots are autonomous. 
\item Robots execute the cycle ({\it Look-Compute-Wait-Move}) asynchronously.
\item Robots are anonymous and homogeneous in the sense that
they are not uniquely identifiable, neither with a unique
identification number nor with some external distinctive mark (e.g., color,
flag, etc.). 
\item A robot can sense all other robots irrespective of their configuration.
\item Each robot is represented as a unit disc (\emph{fat robots}).
\item CORDA \cite{prencipe2001} model is assumed for robots' movement. Under
this model the movement of the robots is not instantaneous. While in motion,
a robot may be observed by other robots.
\item Robots have no common orientation or scale. Each robot uses its own
local coordinate system (origin, orientation and distance). A robot has no
particular knowledge about the coordinate system of any other robot, nor of
a
global coordinate system.
\item Robots can not communicate explicitly. Each robot has a
camera or sensor which can take picture or sense over 360 degree. The robots
communicate only by
means of observing other robots using the camera or sensor. A robot can compute
the coordinates (w.r.t. its own coordinate system) of other robots by observing
through the camera or sensor. 
\item Robots have infinite visibility range.
\item Robots are assumed to be transparent to achieve full visibility.
\item Robots are oblivious. They do not remember the data from the previous
cycles. 
\item Initially all robots are stationary. 
\end{itemize}

Let us define the {\it Gathering Pattern} which is to be formed by $R$. 

The desired gathering pattern for transparent fat robots is shown in
Fig. 1(a). One robot with center $C$ is at the center of the structure. We call
it layer $0$. Robots at layer $1$ are around layer $0$ touching $cir(C,1)$ ($cir(a,n)$ is a circle with center at $a$ and radius $n$).
Robots at layer $2$ are around layer $1$ touching $cir(C,3)$ and so on. Fig.
1(a) shows a gathering pattern with $3$ layers. 

\begin{figure}[!h]
\centering
\includegraphics[height=25mm, width=50mm]{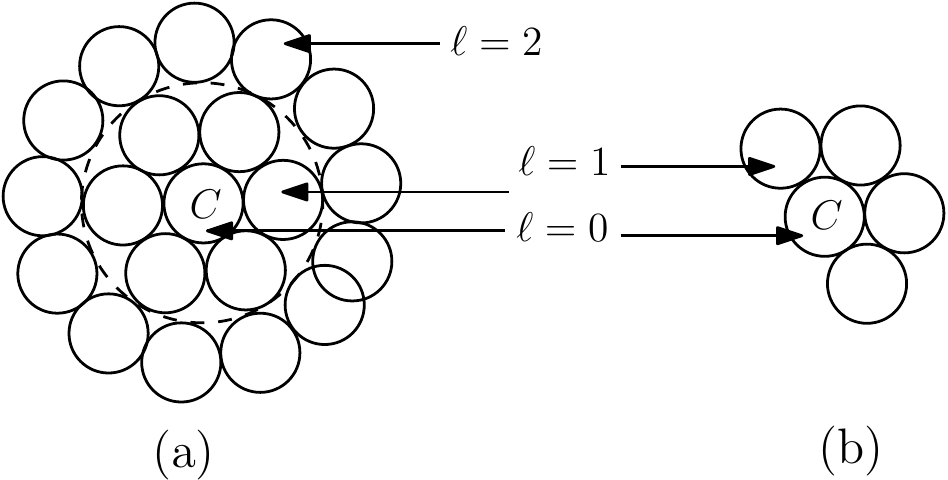}
\caption{{\it Gathering pattern} of multiple robots}
\label{gathpat}
\end{figure}

\begin{defi}
\label{gathering_pattern}
\emph{Gathering pattern} is a circular layered structure of robots with a unique robot
centered at $C$. A robot at layer $\ell+1$ touches $cir(C,2\ell+1)$ and at least
$1$ and at most $2$ robots at layer $\ell+1$. The inner layers are
full. The outer most layer may not be full but the vacant places in 
the outer most layers must be contiguous.  
\end{defi}

According to the definition, a unique center is required to form a
gathering pattern and the layers are created around the center. Finding a
unique robot and marking it as the center is not possible for $2$, $3$ and $4$
robots. This is due to the symmetric structure (Fig. 2). For 2 or 3 robots all
robots are potential candidates to be the center. For 4 robots there are two
such robots. To identify a unique robot for the center of the gathering pattern,
at least $5$ robots are required. For $5$ robots, a robot which touches the rest
$4$ robots is treated as the central robot or a robot at layer $0$. Fig. 1(b)
shows the desired gathering pattern
with minimum robots. 

\begin{figure}[!h]
\centering
\includegraphics[height=10mm, width=40mm]{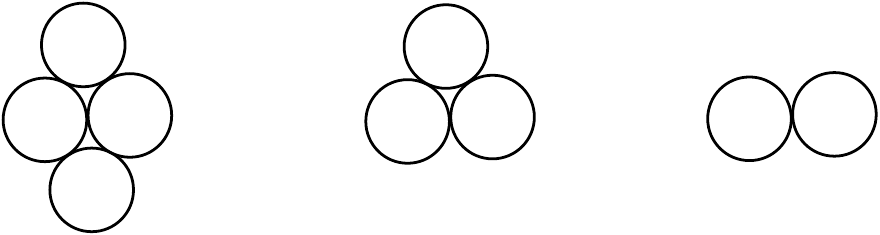}
\caption{Gathering of 2, 3 and 4 robots}
\label{gathpatsym}
\end{figure}

We are given a set, $R$, of separated, stationary transparent fat robots. The
objective is to move the robots so as to build the gathering pattern with the
robots in $R$. For gathering multiple robots, our approach is to select one
robot and assign it
a destination. All other robots remain stationary till the designated robot
reaches its destination. At each turn, the robot selected for motion is chosen
in such a way, that during the motion if any other robot takes a snapshot and
computes, this particular robot would be the only one eligible for movement. To
implement this, we choose a destination carefully. The robot, nearest to the
destination, is allowed to move first. However, a problem will occur
when multiple robots are at same distance from the destination. A leader
election mechanism is required to select a robot among such multiple eligible
robots. To overcome this
problem, our approach is to order a set of robots on 2D plane, with respect to
the destination. The robots move one by one, according to this order, towards the
destination. The mutual ordering of the robots in the set is invariant, though
the size of the set changes.

\section{Characterization of the Geometric Configurations of Robots for
Gathering }
\label{char}

A set of robots on 2D plane is given. Our objective is to move them, one by 
one, to form desire {\it gathering pattern}. 
In order to do so, we create an ordering among the robots and move one by one following the ordering. This
ordering will be computed at different robot sites, who have different
coordinate systems, orientations and origins. Thus, we need the ordering
algorithm to yield the same result, even if the point set is given a
transformation with respect to origin, axes and scale. For this to succeed, 
no two robots should have the same view. Keeping this in mind, we proceed to 
characterize the geometric configurations that is required for a set of robots 
to be orderable (formal definition is presented shortly). In this section we represent the 
robots by points. 

Let $P = \{ p_1, p_2, \ldots, p_n\}$ be a non-empty set of points on the 2D plane. $\cal L$ is a line on 
the plane. Let $P_{\cal L} \subset P$ be the points of $P$ (not all), which lie
on $\cal L$. 
$\cal L$ partitions $P \setminus P_{\cal L}$ into two subsets $H$ and $H'$ where
$H$ or $H'$ 
is non-empty. Let $\cal L'$ be the straight line that intersects $\cal L$ at point
$m$
such that ${\cal L'} \bot {\cal L}$ and  $m$ is the middle point of the span of
the points on $P_{\cal L}$. 

\begin{figure}[!h]
\centering
\includegraphics[height=40mm, width=80mm]{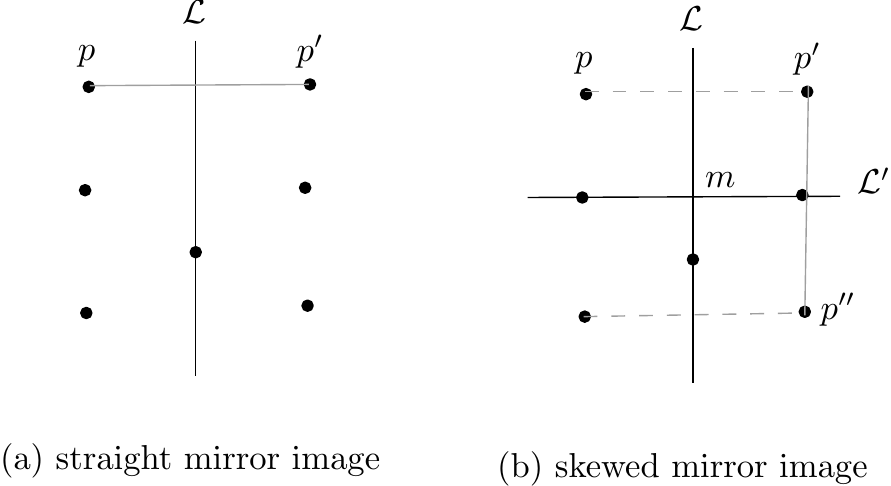}
\caption{Examples of straight and skewed mirror images.}
\label{mirrorimage}
\end{figure}

\begin{defi}
A point $p' \in P$ is said to be a straight mirror image of $p \in P$ across $\cal
L$, if $p'$ is a mirror image of $p$ across $\cal L$ (Figure
\ref{mirrorimage}(a)). 
\label{stmirrorimage}
\end{defi}

\begin{defi}
Let $p'$ be the straight mirror image of $p$ across $\cal L$. $p''$ is the
straight mirror image of $p'$ across $\cal L'$. $p''$ is said to be skewed
mirror image of $p$ across $\cal L$ (Figure
\ref{mirrorimage}(b)).
\label{skmirrorimage}
\end{defi}

\begin{figure}[!h]
\centering
\includegraphics[height=40mm, width=85mm]{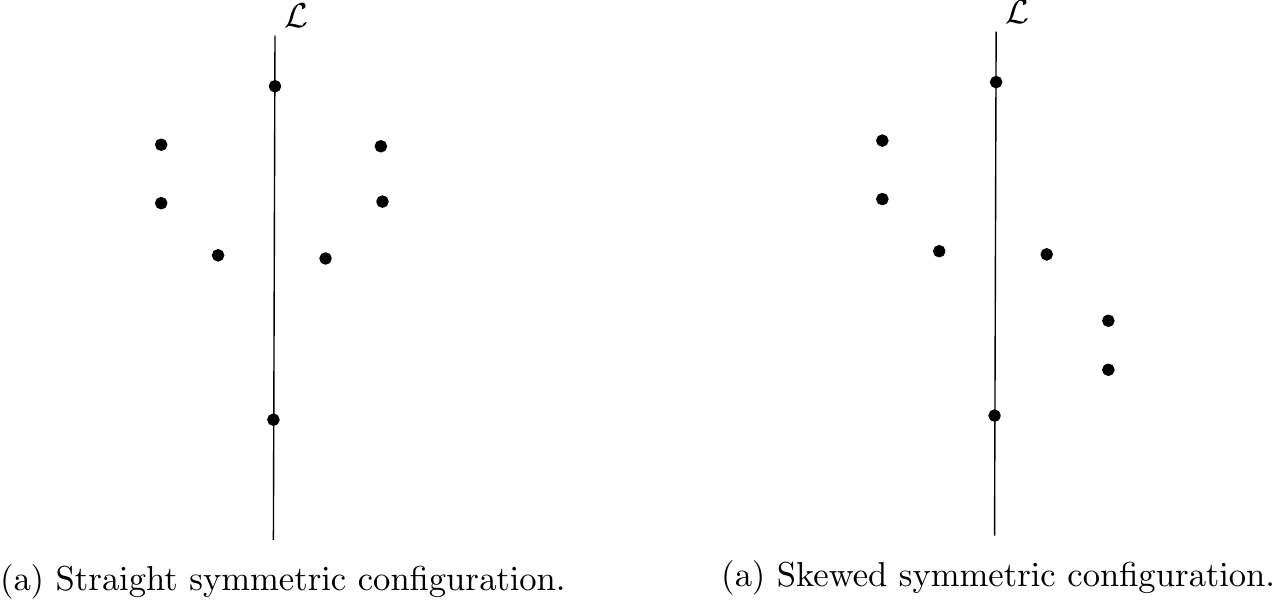}
\caption{Examples of straight and skew symmetric configurations.}
\label{symconfig}
\end{figure}

\begin{defi}
A set of points $P$ on the 2D plane is said to be in straight-symmetric
configuration, if there exists a
straight line $\cal L$ (on that
plane) not containing all the points of $P$, such that each point in $H \cup
P_{\cal L}$ has a straight mirror image in  $H' \cup P_{\cal L}$ (Figure. \ref{symconfig}(a)).
The line $\cal L$ is called a line of straight symmetry.
\label{def-straight-symmetric-conf}
\end{defi} 

Note that each point in $P_{\cal L}$ is the mirror image of itself. 

\begin{defi}
A set of points $P$ on the 2D plane is said to be in skew-symmetric
configuration, if there exists a
straight line $\cal L$ (on that
plane) not containing all the points of $P$, such that each point in $H \cup
P_{\cal L}$ has a skewed mirror image in  $H' \cup P_{\cal L}$. (Figure. \ref{symconfig}(b)).
The line $\cal L$ is called a line of skew symmetry.
\label{def-skewed-symmetric-conf}
\end{defi}

\begin{defi}
\label{def-symmetric-conf}
A set of points $P$ on the 2D plane is said to be in symmetric
configuration (denoted by $\O{}_S$), if it is either a singleton 
set or in straight symmetric or skew symmetric configuration. For 
a singleton set, any line passing through the point is a line of 
symmetry. 
\end{defi}

\begin{defi}
\label{def-asymmetric-conf}
A set of points $P$, which is not in symmetric configuration is in
asymmetric configuration (denoted by $\O{}_A$).
\end{defi}

Our requirement does not stop at requiring the algorithm to be robust to changes
in the coordinate system. The positions of the robots also change as the
algorithm progresses. Our objective is to order the points (robots) in $P$ such
that when the robots move one by one, according to this order, towards the
destination, the mutual ordering of the robots in the set is invariant. We
define an orderable set as follows.

\begin{defi}
\label{def-orderable-set}
A set of points $P$, on the plane, is called an orderable set, if there
exists a deterministic algorithm, which produces a unique ordering of the
points of $P$, such that the ordering is same irrespective of the choice of
origin and coordinate system.
\end{defi}

\begin{lemma}
\label{lemma-sym-ord}
Let $P$ be a non-empty, non-singleton set of points. If $P$ is in $\O{}_s$, then $P$ is not orderable.
\end{lemma}

\emph{Proof:}
Let $\cal L$ be a line of symmetry (straight or skewed) for $P$. Let $P_{\cal
L}$ be
the set of points from $P$, lying on $\cal L$. $\cal L$ divides $P \setminus
P_{\cal L}$ into two halves $H$ and $H'$. $H \cup P_{\cal L}$ and $H'\cup
P_{\cal L}$ are mirror images (straight or skewed) 
of each other. Let $p_i$ be a point in $H$ and $p'_i$  in $H'$, the mirror image
of $p_i$. Consider an arbitrary ordering algorithm $\cal A$. If we run $\cal A$
on $P$ with $p_i$ as the origin, it produces an ordering of $P$. Let $p_j$ be
the first point from that ordering such that $p_j$ is not on $P_{\cal L}$. On
the other hand, if we run $\cal A$ on $P$, with $p'_i$ as the
origin, the symmetry tells us that the ordering obtained will have $p'_j$ (the
mirror image of $p_j$) as the corresponding first point in the order. Since,
$p_j$ is
not in $P_{\cal L}$, $p_j \ne p'_j$. Since the choice of $\cal A$ was arbitrary, no
algorithm will produce the same order irrespective of the choice of origin.
Hence, $P$ is not orderable. 
\qed

\begin{obs}
\label{lemma-LOS-Passes-CenterSECP}
Any line of symmetry of $P$ passes through the center of the Smallest Enclosing Circle (SEC) of
$P$ and divides the points on SEC into two equal mirror images (straight or
skewed). 
\end{obs}

In order to check whether a set of points is in $\O{}_s$ or not, we need to find out if
a line of symmetry exists for this set. For this search to be feasible, we need to
reduce the potential set of candidate lines for line of symmetry. In order to do so, first the SEC of $P$ is computed. The points on the SEC are taken to form a convex polygon, say ${\cal H}(P)$.   

\begin{obs}
A line of symmetry (straight or skewed) of $P$ cuts ${\cal H}(P)$ at two points.
Thus the line of symmetry contains at most two points from ${\cal H}(P)$. 
\end{obs}

\begin{lemma} \label{hull-vertex-symmetry}
Let $P$ be a set of points in $\O{}_s$. The mirror image (straight or skewed)
of a vertex in $\cal H$ is also in ${\cal H}(P)$.
\end{lemma} 

\begin{figure}[!h]
\centering
\includegraphics[height=45mm, width=75mm]{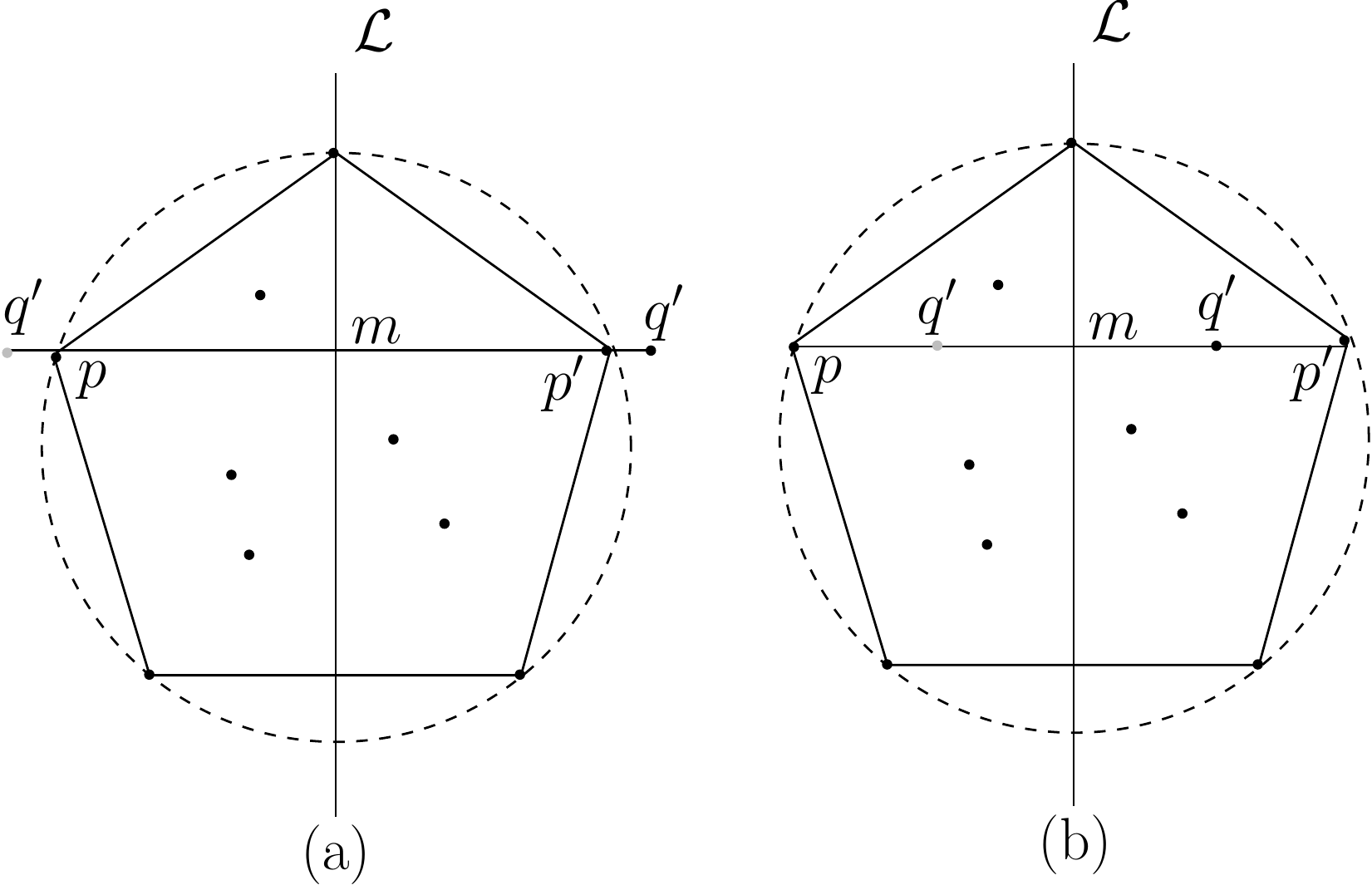}
\caption{The mirror image of a hull vertex is also a hull vertex.}
\label{hullimage}
\end{figure}

\emph{Proof:}
Consider a vertex $p$ of ${\cal H}(P)$. Let $p'$ be the mirror image of $p$ across the line of symmetry $\cal L$. Suppose, $p'$ is not in ${\cal
H}(P)$.  Suppose,
the line $pp'$ intersects ${\cal L}$ at $m$ and intersects ${\cal H}(P)$ at
$q$. $\overline{pm} = \overline{mp'}$ \footnote{ If $a$ and $b$ are two points on the 2D plane then the distance between $a$ and $b$ is represented by $\overline{ab}$}. $q$ should also have a mirror image across $\cal L$, in the
same side where $p$ lies. Let $q'$ be the mirror image of $q$. Note that $qq'$
also intersects $\cal L$ at point $m$  and $\overline{qm} = \overline{mq'}$. If $p'$ is inside
the convex hull then $\overline{mp'} < \overline{mq}$ (Figure. \ref{hullimage}(a)). Hence, $\overline{mp} < \overline{mq}$
and $\overline{mp} < \overline{mq'}$. This implies that $p$ is not a hull vertex. Contradiction! On the other hand, if $p'$ is outside the convex 
hull then $\overline{mp'} > \overline{mq}$ (Figure. \ref{hullimage}(b)). Hence, $\overline{mp} > \overline{mq}$ and $\overline{mp} > \overline{mq'}$. 
This implies that $p$ is not a hull vertex. Contradiction! Therefore,
if $p$ is a hull vertex, $p'$ must also be a hull vertex.
\qed

\begin{lemma}
If $P$ is in $\O{}_s$, then ${\cal H}(P)$ is in $\O{}_s$.  
\end{lemma}

\emph{Proof:}
Follows from lemma \ref{hull-vertex-symmetry}.
\qed

\begin{lemma}
\label{LemmmaStraightSymHull}
If ${\cal H}(P)$ is in $\O{}_S$, then for any line of straight symmetry $\cal L$
of $P$,
\begin{enumerate}
\item if $\cal L$ passes through a vertex $v$ of ${\cal H}(P)$, it bisects the
interior angle at $v$,
\item if $\cal L$ intersects an edge $e$ of ${\cal H}(P)$, at a point other than
a
vertex, it is the perpendicular bisector of $e$.
\end{enumerate}
\end{lemma}

\emph{Proof:} Follows from the proof of lemma \ref{hull-vertex-symmetry}.
\qed

\begin{obs}
\label{LoSkS-intersect-1-vertex-edge}
Any line of skew symmetry  intersects ${\cal H}(P)$ either at two vertices or at two edges.
\end{obs}

\begin{obs}
\label{LoSkS-intersect-2-vertex-edge}
A skew symmetric polygon has even number of vertices and edges.
\end{obs}

\begin{lemma}
\label{lem-rectangle}
A pair of edges in a polygon inscribed in a circle is parallel and equal if and only if they are opposite sides of a unique rectangle inscribed in that circle.
\end{lemma}

\begin{figure}[!h]
\centering
\includegraphics[height=45mm, width=40mm]{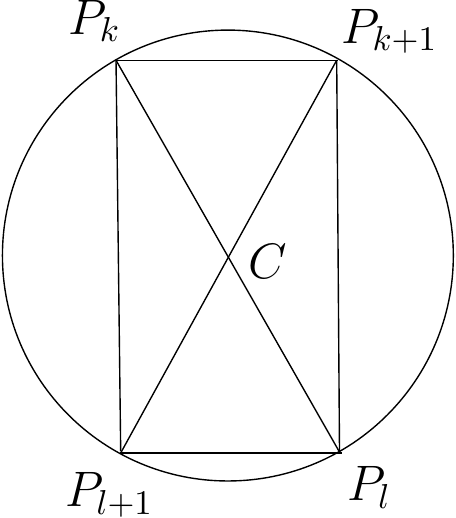}
\caption{A pair of parallel and equal edges of a polygon forms a rectangle.}
\label{skew1}
\end{figure}

\emph{Proof:} 
    
{\bf If:} Trivial. 

{\bf Only if:} Suppose $P_{k}P_{k+1}$ and $P_{l}P_{l+1}$ are two edges of a polygon such that $\overline{P_{k}P_{k+1}} =\overline{P_{l}P_{l+1}}$ and $P_{k}P_{k+1} || P_{l}P_{l+1}$ (Fig. \ref{skew1}). 
$C$ is the intersection point of the lines $P_kP_l$ and $P_{k+1}P_{l+1}$. 
It is easy to see that $\triangle P_{k}CP_{k+1} \cong \triangle P_{l}CP_{l+1}$. So, $\overline{P_{k+1}C} = \overline{P_{l+1}C}$ and $\overline{P_{k}C} = \overline{P_{l}C}$. This means that the cords, $P_kP_l$ and $P_{k+1}P_{l+1}$ bisect each other. Hence, $P_kP_l$ and $P_{k+1}P_{l+1}$ are both diameters of the circle. Therefore, $\overline{P_kP_l} = \overline{P_{k+1}P_{l+1}}$. 
This implies that $P_kP_{k+1}P_{l}P_{l+1}$ is a rectangle.
\qed 

\begin{figure}[!h]
\centering
\includegraphics[height=45mm, width=50mm]{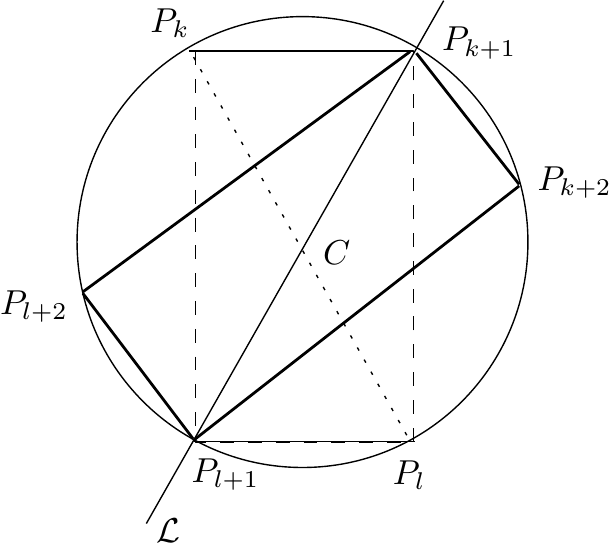}
\caption{A pair of parallel and equal edges of a polygon forms rectangles.}
\label{skew}
\end{figure}
\begin{lemma}

\label{obs-parallel-edge-equal}
A polygon ${\cal H}(P)$, inscribed in a circle, is skew symmetric if and only if each edge of the polygon has a parallel edge of equal length.
\end{lemma}

\emph{Proof:}
{\bf If:} 
Suppose $P_{k}P_{k+1}$ and $P_lP_{l+1}$, are parallel and equal edges of ${\cal H}(P)$ (Fig \ref{skew}). 
Let us add $P_{k+1}$ and $P_{l+1}$ by a straight line $\cal L$. We shall show that $\cal L$ is a line of skew symmetry for ${\cal H}(P)$. $P_{l+1}P_{l}$ is the skewed mirror image of $P_kP_{k+1}$ across $\cal L$. Let $P_{k+1}P_{k+2}$ be the adjacent edge of $P_kP_{k+1}$. We add $P_{k+2}P_{l+1}$. Since, $P_{k+1}P_{l+1}$ is a diameter of the circumcircle of ${\cal H}(P)$ (lemma \ref{lem-rectangle}), $\angle P_{k+1}P_{k+2}P_{l+1} = 90$ degree. We draw the rectangle $P_{k+1}P_{k+2}P_{l+1}P_{l+2}$. By lemma \ref{lem-rectangle}, $P_{l+1}P_{l+2}$ is the edge of the polygon which is parallel to and equal in length with $P_{k+1}P_{k+2}$. By repeating this argument it can be shown that, polygonal chains on both sides of $\cal L$ are skew symmetric. 

{\bf Only if:} 
Let  ${\cal H}(P)$ be a skew symmetric polygon. $\cal L$ is a line of skew symmetry for ${\cal H}(P)$. $\cal L$ partitions  ${\cal H}(P)$ into two halves namely, $H_1$ and $H_2$.  

\begin{figure}[!h]
\centering
\includegraphics[height=40mm, width=65mm]{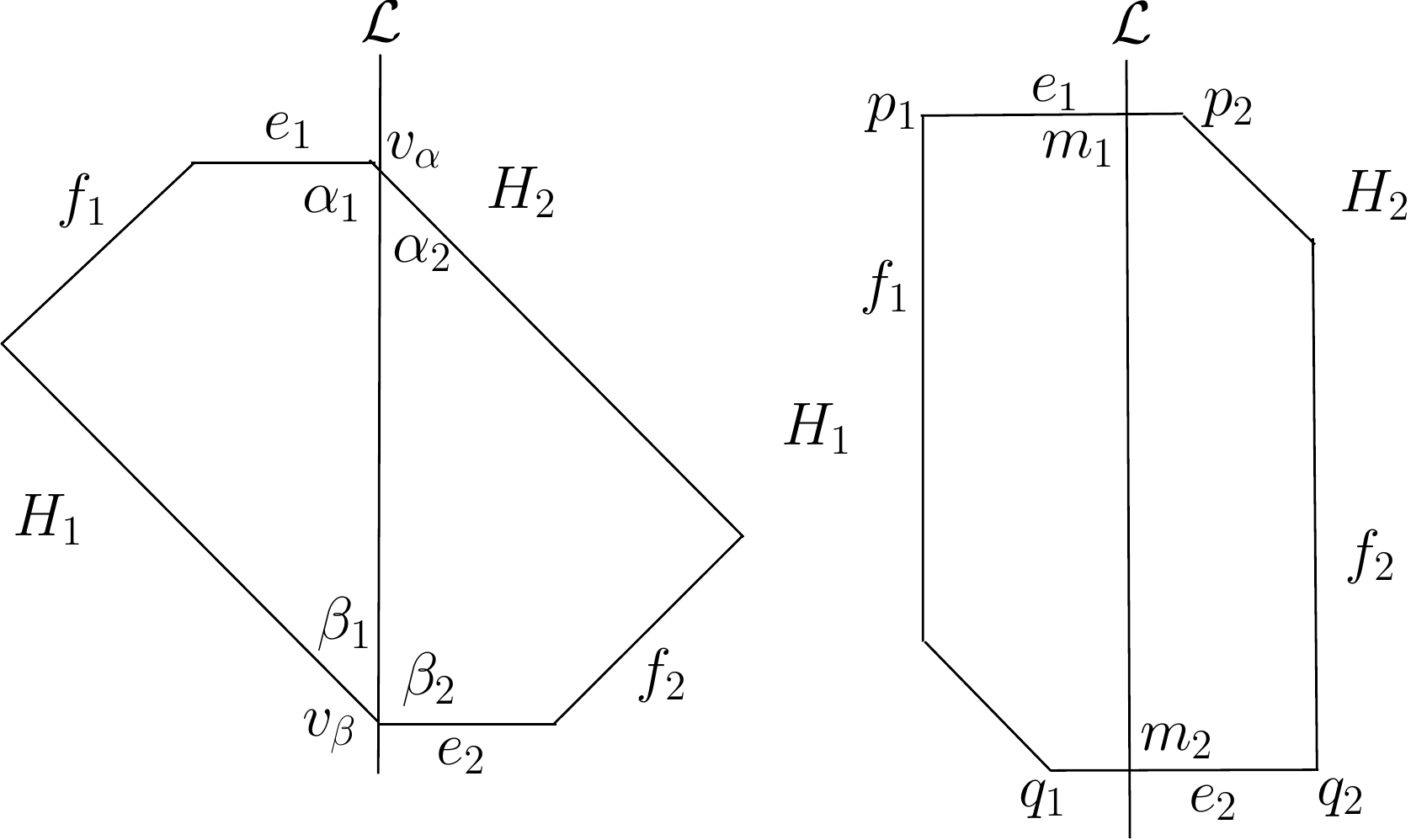}
\caption{Examples of skew symmetry in convex hull.}
\label{SkewSymHFig}
\end{figure}

First consider the case when $\cal L$ passes through two vertices of  ${\cal H}(P)$,  namely
$v_{\alpha}$ and $v_\beta$. (Fig \ref{SkewSymHFig}). Suppose, $e_1$ is the edge incident at $v_{\alpha}$ in $H_1$ and 
$e_2$ is the edge incident at $v_{\beta}$ in $H_2$. Since, $\cal L$ is a line of skew symmetry, $e_2$ is the skewed mirror image of $e_1$. 
Therefore, $\overline{e_1} = \overline{e_2}$ \footnote{The length of the edge $e$ is denoted by $\overline{e}$} and $e_1 || e_2$. Let $f_1$ be the edge adjacent to $e_1$ in $H_1$ and $f_2$ the edge adjacent to $e_2$ in $H_2$.
Similarly, $f_2$ is the mirror image of $f_1$. Therefore, $\overline{f_1} = \overline{f_2}$ and $f_1 || f_2$. In this manner, we can find a parallel and equal edge of every edge. 

Now consider the case when $\cal L$ intersects two edges of ${\cal H}(P)$, namely $e_1 =
p_1p_2$ and $e_2 = q_1q_2$ at point $m_1$ and $m_2$ respectively (Fig \ref{SkewSymHFig}).
If we consider a modified polygon with additional vertices at $m_1$ and $m_2$, the result follows from the previous case.  
\qed

Suppose, ${\cal H}(P)$ is a skew symmetric polygon. Let $\cal L$ be a line intersecting two vertices of ${\cal H}(P)$, namely
$v_{\alpha}$ and $v_\beta$. Let $\alpha$ be the interior angle of ${\cal H}(P)$ at vertex
$v_{\alpha}$ and $\beta$ be the interior angle of ${\cal H}(P)$ at vertex
$v_{\beta}$. $\cal L$ divides $\alpha$ into ${\alpha}_1$ and ${\alpha}_2$
and $\beta$ into ${\beta}_1$ and ${\beta}_2$ (Fig \ref{SkewSymHFig}). Suppose, $e_1$ is an edge incident at $v_{\alpha}$ in $H_1$ and 
$e_2$ is an edge incident at $v_{\beta}$ in $H_2$. $f_1$ is the edge adjacent to $e_1$ in $H_1$ and $f_2$ is the edge adjacent to $e_2$ in $H_2$.   

\begin{lemma} \label{LoSkS-vertex}
For a skew symmetric polygon ${\cal H}(P)$, $\cal L$ is a line of skew symmetry if and only if ${\alpha}_1 = {\beta}_2$ or ${\alpha}_2 = {\beta}_1$.
\end{lemma}

\emph {Proof:}
{\bf If:} If ${\alpha}_1 = {\beta}_2$, $e_1 || e_2$. From lemma \ref{obs-parallel-edge-equal}, there exist an edge $e_1'$ such that $e_1 || e_1'$
and $\overline{e_1} = \overline{e_1'}$. As ${\cal H}(P)$ is convex $e_1 || e_1'$ and $e_1 || e_2$ implies $e_1'$ and  $e_2$ are the same. 
Hence, $\overline{e_1} = \overline{e_2}$. Using an argument similar to that used in the proof of lemma \ref{lem-rectangle}, it 
can be shown that, polygonal chains on both sides of $\cal L$ are skew symmetric. 
Similarly, if ${\alpha}_2 = {\beta}_1$, polygonal chains on both sides of $\cal L$ are skew symmetric.  
Hence, $\cal L$ is a line of skew symmetry.

{\bf Only if:} Follows from the definition of skew symmetric polygon. 
\qed

Let $\cal L$ be a line intersecting two edges of ${\cal H}(P)$, namely $e_1 =
p_1p_2$ and $e_2 = q_1q_2$ at point $m_1$ and $m_2$ respectively (Fig \ref{SkewSymHFig}).

\begin{lemma} \label{LoSkS-edge}
For a skew symmetric polygon ${\cal H}(P)$,
$\cal L$ is a line of skew symmetry if and only if
$e_1$ $||$ $e_2$, $\overline{p_1m_1} = \overline{m_2q_2}$ and $\overline{m_1p_2} = \overline{q_1m_2}$.
\end{lemma} 

\emph{Proof:}
{\bf If:} $e_1$ $||$ $e_2$, $\overline{p_1m_1} = \overline{m_2q_2}$ and $\overline{m_1p_2} = \overline{q_1m_2}$ implies that $q_2$ 
is the skewed mirror image of $p_1$ and $p_2$ is the skewed mirror image of $q_1$ across $\cal L$. As $\overline{e_1} = \overline{e_2}$, using an 
argument similar to that used in the proof of lemma \ref{lem-rectangle}, it can be shown that, polygonal chains on both sides of $\cal L$ 
are skew symmetric. Hence, $\cal L$ is a line of skew symmetry. 

{\bf Only if:} Follows from the definition of skew symmetric polygon.       
\qed

In order to check whether a set of points $P$ is in $\O{}_s$ or not, 
we first compute the SEC of $P$. The convex polygon ${\cal H}(P)$, as described earlier, is also computed. For each vertex 
and each edge of ${\cal H}(P)$, we look for a line of  symmetry $\cal L$
(straight or skewed) passing through that vertex or edge. 

Since, we want the ordering to be the same for any choice of origin and axes, we
can only use information which are invariant under these transformations.
Examples of such properties are, distances and angles between the points. The distances may be
affected by the choice of unit distance, but even then their ratios remain the
same. One possible solution is to select the robot closest to the destination as
the leader or the candidate to move. It also satisfies our extra requirement
that it remains the point closest the destination, and hence the leader, as it
moves. Robots equidistant from the destination are on the circumference of a
circle with the destination as its center. They also form a convex polygon. Let
$G$ be a set of robots forming such a convex polygon inscribed in a circle. 
For the rest of this paper, by {\it convex polygon} we mean such polygons which are inscribed in a circle. 
The center of the circle is called the center of the polygon.

\begin{defi}
\label{def-straight-symmetric-poly}
A convex polygon $G$ is {\it straight symmetric} if the set of vertices in $G$
is in straight symmetric configuration.
\end{defi}
\begin{defi}
\label{def-skew-symmetric-poly}
A convex polygon $G$ is {\it skew symmetric} if the set of vertices in $G$
is in skew symmetric configuration.
\end{defi}
\begin{defi}
\label{def-symmetric-poly}
A convex polygon $G$ is symmetric if it is either straight symmetric or
skew symmetric.
\end{defi}
\begin{defi}
\label{def-asymmetric-poly}
A polygon, which is not  symmetric is asymmetric. 
\end{defi}
   
Note: A single point on a circle is a special case. It is symmetric. Though, any line passing through it is a line of symmetry, we shall only call the line passing through the center of the circle (and the point itself) as the line of symmetry.

\begin{obs}
\label{obs-asym-poly}
The set of vertices of an asymmetric polygon is in $\O{}_A$. 
\end{obs}

\begin{figure}[!h]
\centering
\includegraphics[height=40mm, width=30mm]{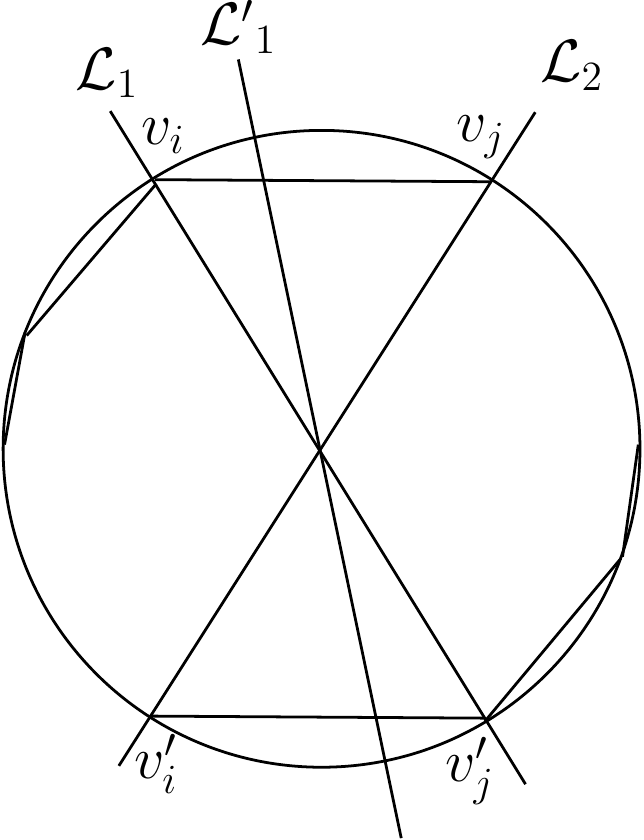}
\caption{Any straight line passing through the center of a skew symmetric polygon, is a line of skew symmetry
for that polygon}
\label{skew2}
\end{figure}

\begin{lemma}
\label{lem-skew-los}
Any straight line passing through the center of a skew symmetric polygon, is a line of skew symmetry
for that polygon.   
\end{lemma}
\emph{Proof:}
Let $G$ be a skew symmetric polygon. Let $v_iv_j$ and $v'_iv'_j$ be two parallel and equal length edges of $G$ (Figure. \ref{skew2}). 
Let ${\cal L}_1$ and ${\cal L}_2$ be the lines passing through $v_i$-$v'_j$ and $v_j$-$v'_i$ respectively.
From lemma \ref{lem-rectangle}, it follows that ${\cal L}_1$ and ${\cal L}_2$ pass through the center $C$ of $G$ and they are lines of skew symmetry of $G$.
Now it is sufficient to prove that any line passing through the center and intersecting $\overline{v_iv_j}$ is a line of skew symmetry for $G$.
Without loss of generality let us rotate ${\cal L}_1$ by some angle, around $C$, keeping it between ${\cal L}_1$ and ${\cal L}_2$. Let ${\cal L}'_1$ be the new position of the line. Following lemma \ref{LoSkS-edge} and 
using an argument similar to that used in the proof of lemma \ref{lem-rectangle}, it can be shown that, polygonal chains on both sides of ${\cal L}'_1$ 
are skew symmetric. Hence, ${\cal L}'_1$ is a line of skew symmetry.         
\qed
   
\begin{figure}[!h]
\centering
\includegraphics[height=50mm, width=120mm]{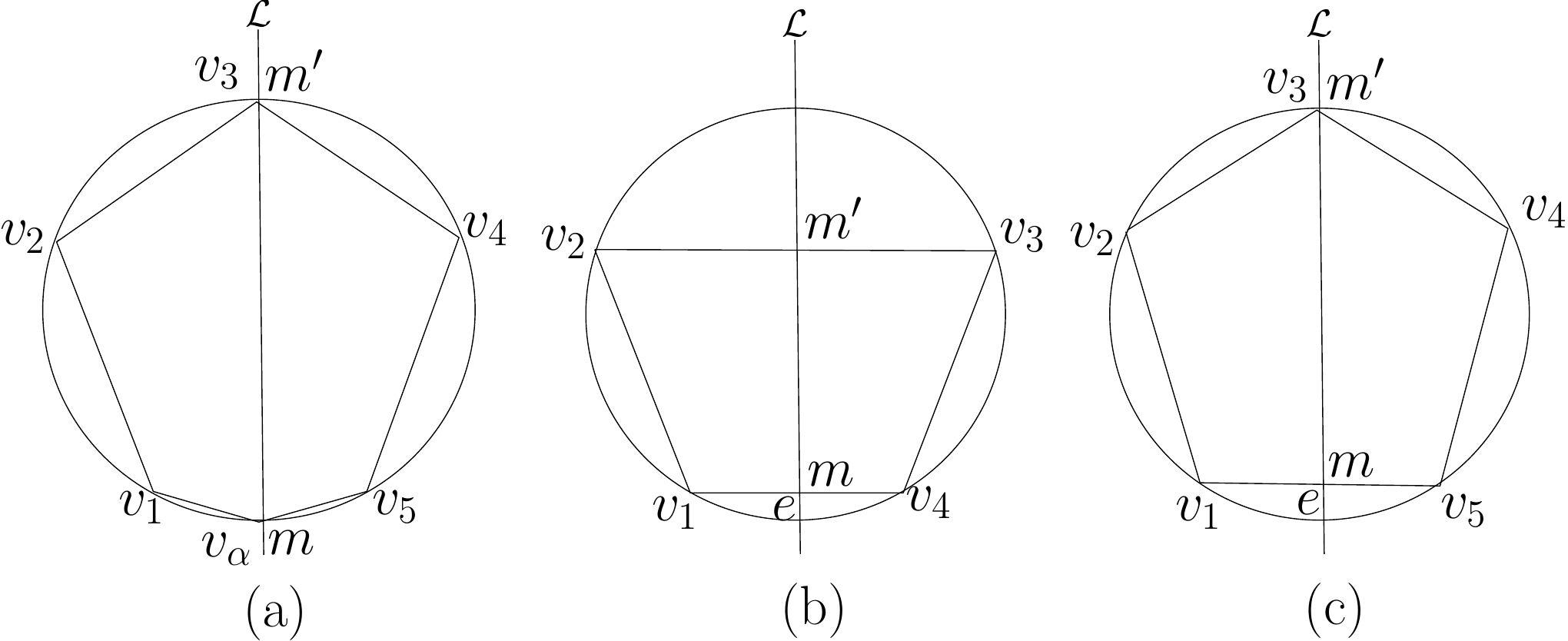}
\caption{Examples of symmetric polygons with line of symmetry intersecting the
polygon by two vertices (a) and by two
edges (b) and by one edge and one vertex (c).}
\label{symmetric}
\end{figure}

For a convex polygon, a line of symmetry $\cal L$ (straight or skewed) 
intersects the 
polygon at two points. The
points can be two vertices or they may lie on two edges or one point may be a
vertex and the other point lies on an edge (Figure. \ref{symmetric}). Let $\cal
L$ intersect $G$ at $m$
and
$m'$. Suppose $G$ has $k$ vertices. The vertices are labeled starting from the
vertex next (clockwise) to $m$ up to the previous vertex of $m$, as $v_1, v_2,
\ldots, v_n$. If $m$ is a vertex, then $n = k-1$. If $m$ lies on an
edge, $n = k$. Similarly, the vertices are labeled starting from the vertex
next (clockwise) to $m'$ up to the previous vertex of $m'$, as $v'_1, v'_2,
\ldots, v'_{n'}$. If $m'$ is a vertex, then $n' = k-1$. If $m'$ lies on an
edge, $n' = k$.

The following notations are used in the rest of this paper. 
\begin{itemize}
\item $CW_{m} = (\overline{mv_1},\overline{mv_2},\ldots,\overline{mv_n})$.
\item $ACW_{m} = (\overline{mv_n},\overline{mv_{n-1}},\ldots,\overline{mv_1})$.
\item $CW_{m'} = (\overline{m'v'_1},\overline{m'v'_2},\ldots,\overline{m'v'_{n'}})$.
\item $ACW_{m'} = (\overline{m'v'_{n'}},\overline{m'v'_{n'-1}},\ldots,\overline{m'v'_1})$.
\end{itemize}

$CW_{m} = ACW_{m}$ iff $\overline{mv_i} = \overline{mv_{n-i+1}}$ for ($1 \le i \le n$).

\begin{lemma}
\label{lemma-straight-sym}
Let $G$ be a straight symmetric polygon. A straight line $\cal L$ is a line of straight
symmetry for $G$ if and only if $CW_{m} = ACW_{m}$ or $CW_{m'} = ACW_{m'}$.
\end{lemma}

\emph{Proof:}
Follows from definition \ref{def-straight-symmetric-conf} and definition \ref{def-straight-symmetric-poly}.
\qed

\begin{lemma}
\label{lemma-skewed-sym}
Let $G$ be a skew symmetric polygon and $\cal L$ a line of skew
symmetry. $\cal L$ divides $G$ into two skewed mirror image parts if and only if
$CW_{m} = CW_{m'}$ or $ACW_{m} = ACW_{m'}$.
\end{lemma}

\emph{Proof:}
Follows from definition \ref{def-skewed-symmetric-conf} and definition \ref{def-skew-symmetric-poly}.
\qed

\begin{theorem}
\label{obs-asymmetric-polygon}
A polygon $G$ is asymmetric if and only if all of the following conditions are
true
\begin{enumerate}
\item $CW_{v_i} \ne ACW_{v_i}$ for $1 \le i \le n$.
\item $CW_{v_i} \ne CW_{v_j}$  for $1 \le i, j \le n$ and $i \ne j$.
\item $CW_{v_i} \ne ACW_{v_j}$  for $1 \le i, j \le n$ and $i \ne j$
\end{enumerate}
\end{theorem}

Note: $CW_{v_i} \ne CW_{v_j}$ and $ACW_{v_i} \ne ACW_{v_j}$ are equivalent.
Theorem 1 can also be written as follows.

{\bf Theorem 1.1}: 
A polygon $G$ is asymmetric if and only if all of the following conditions are
true
\begin{enumerate}
\item $CW_{v_i} \ne ACW_{v_i}$ for $1 \le i \le n$.
\item $ACW_{v_i} \ne ACW_{v_j}$   for $1 \le i, j \le n$ and $i \ne j$.
\item $ACW_{v_i} \ne CW_{v_j}$.  for $1 \le i, j \le n$ and $i \ne j$
\end{enumerate}
  
\begin{lemma}
\label{lem-asym-ord}
An asymmetric polygon is orderable.
\end{lemma}
\emph{Proof:} Let $G$ be an asymmetric polygon. For each vertex $v_k$
($1 \le k \le n$), we compute the tuple $\{CW_{v_k}, ACW_{v_k}\}$ and take the lexicographic minimum of the $n$ tuples as the ordering of $G$. 
Since $G$ is asymmetric, $CW_{v_i} \ne ACW_{v_i}$ for any $v_i$ and $CW_{v_i} \ne CW_{v_j}$, $ACW_{v_i} \ne
ACW_{v_j}$ for any $v_i$ and $v_j$ such that $i \ne j$ (theorem \ref{obs-asymmetric-polygon}).
Hence, the ordering is same irrespective of the choice of origin or the coordinate axes.
Thus, the set of vertices of an asymmetric polygon is orderable (definition \ref{def-orderable-set}).
\qed

A set of points which are equidistant from the destination, form a convex polygon such that the vertices of the polygon are on
the circumference of a circle. 
There can be multiple such sets, i.e.,
the robots within a set are equidistant from the destination but the
robots in different sets are not equidistant from the destination. 
In such a case, we get multiple concentric circles each of them enclosing a
convex polygon.

Let $\cal G$ = $(G_1, G_2, \ldots, G_t)$, be the $t$ convex polygons
whose union is the whole set of points. The vertices of each polygon in $\cal G$
are on the circumference of a circle. 
These circles are concentric. The center of the circles is known. Generally, the
center $C$ is the center of the SEC formed 
by the points in $\cal G$. $C$ is also considered as the center of any polygon $G_i \in \cal G$. 
Elements of $\cal G$ are sorted according to their distances from the center. Any polygon 
$G_i \in \cal G$ can be extracted by selecting the equidistant points from the
center. The polygon at level $1$ ($G_1$) is the closest and the polygon at level $t$
($G_t$) is the farthest. 

\begin{defi}
\label{def-sym-pair}
A pair of polygons $G_i$ and $G_j$(denoted by $<G_i, G_j>$) in $\cal G$ is
called a \emph{symmetric pair}, if $G_i$ and $G_j$ have a
common line of symmetry. 

A pair is \emph{asymmetric} if it is not symmetric.
\end{defi}

\begin{obs}
\label{obs-pair-asym-ord}
If any of the polygons of a pair is asymmetric, then the pair is asymmetric.
\end{obs}

\begin{lemma}
\label{lem-sym-notord}
A symmetric pair is not orderable.
\end{lemma}

\emph{Proof:}
Suppose $<G_i, G_j>$ is a symmetric pair. $G_i$ and $G_j$ have a common line 
of symmetry (straight or skewed) $\cal L$, which divides $G_i$ and $G_j$ into
two equal 
halves (straight or skewed mirror image). The
union of the vertices of $G_i$ and $G_j$ is divided into two equal halves by $\cal
L$. The union of the vertices of $G_i$ and
$G_j$ is in $\O{}_S$. Hence, $<G_i, G_j>$ is not
orderable (lemma \ref{lemma-sym-ord}).
\qed

Suppose, the vertices of $G_i$ in pair $<G_i, G_j>$ are projected radially on the circumference of the enclosing circle of $G_j$ ($i < j$). 
Construct a polygon $G_{ij}$ with the full set of vertices on the circle.

\begin{obs}
\label{obs-merg-asym-pair}
If $<G_i, G_j>$ is an asymmetric pair, then the polygon $G_{ij}$ is asymmetric.
\end{obs}

\begin{lemma}
\label{lem-asym-pair-ord}
An asymmetric pair is orderable.
\end{lemma}

\emph{Proof:} Let $<G_i,G_j>$ be an asymmetric pair. $G_{ij}$ is an
asymmetric polygon (observation \ref{obs-merg-asym-pair}). Therefore
$G_{ij}$ is orderable (lemma \ref{lem-asym-ord}). Therefore $<G_i,
G_j>$ is orderable.
\qed

\begin{defi}
\label{def-sym-calG} 
$\cal G$ is called symmetric (or $\cal G$ is in $\O_{S}$) if all polygons in
$\cal G$ have a common line of symmetry. 
\end{defi}

\begin{lemma}
\label{lem-all-pair-asym-ord}
If there exists an asymmetric pair in $\cal G$, then $\cal G$ is orderable. 
\end{lemma}

\emph{Proof:}
Let $<G_i,G_j>$ in $\cal G$ be the asymmetric pair such that $(i,j)$ is lexicographically minimum
among all pairs in $\cal G$ which are asymmetric. If $\cal G$ has one pair, i.e., $\cal G$ = $\{G_1,
G_2 \}$ and $<G_1, G_2>$ is asymmetric, then $\cal G$ is orderable (lemma \ref{lem-asym-pair-ord}).

Consider the case when $\cal G$ has more than one pair. Since $<G_i,
G_j>$ is asymmetric, it is orderable (lemma
\ref{lem-asym-pair-ord}).
Let $O_i$ and $O_j$ be two orderings of $G_i$ and $G_j$ respectively. 
Let $G_k$ ($k \ne i, j$) be any polygon in $\cal G$. Since, $<G_i,G_j>$ is
asymmetric, $G_{ij}$
is asymmetric (observation \ref{obs-merg-asym-pair}). Hence, $<G_k, G_{ij}>$ is
asymmetric and is orderable (lemma \ref{lem-asym-pair-ord}). Let $O_k$ be an
ordering of $G_k$. Varying
$k$ from $0$ to $t$ ($k \ne i, j$), we get the ordering of polygons $G_0,
G_1, \ldots, G_t$ in terms of $O_0, O_1, \ldots, O_t$ respectively. Hence, we
get the ordering of $\cal G$.  
\qed

\begin{lemma}
\label{lem-one-pair-asym-ord}
If $\cal G$ contains at least one asymmetric polygon then $\cal G$ is orderable. 
\end{lemma}

\emph{Proof:}
Let $G_a \in \cal G$ be asymmetric. 
For all $G_i \in {\cal G} \setminus G_a$, $<G_i, G_a>$ is asymmetric (observation
\ref{obs-pair-asym-ord}). Hence, following lemma
\ref{lem-all-pair-asym-ord}, $\cal G$ is orderable.
\qed

Let $G_i$, $G_j$, and $G_k$ in $\cal G$ be pairwise symmetric. Let $L_{ij}$,
$L_{jk}$, $L_{kl}$ be the 
lines of (same) symmetry of $<G_i, G_j>$, $<G_j, G_k>$ and 
$<G_k, G_l>$ respectively. Our aim is to show that $G_i$ , $G_j$ and $G_k$ have
a common line of symmetry. 
To show this, we first characterize the polygon on the basis of symmetry they
have. The following lemma states a very interesting property for a convex regular
polygon.

\begin {lemma}
\label{lemma-sym-regular}
If a polygon $G$  is convex and
has more than one line of symmetry such that 
each line of symmetry passes through at least one vertex of $G$,
then $G$ is regular.
\end {lemma}

\begin{figure}[!h]
\centering
\includegraphics[height=60mm, width=50mm]{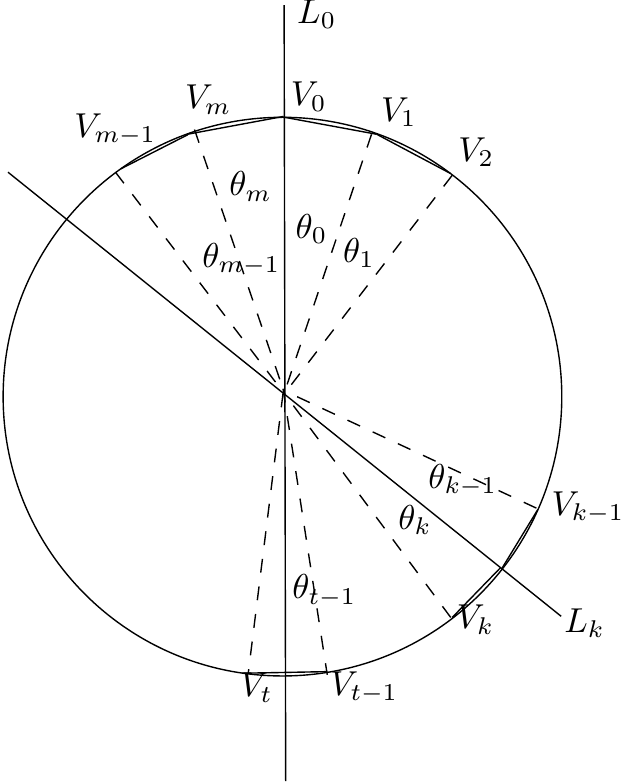}
\caption{An example of a regular convex polygon with lines of symmetry $L_0$
and $L_k$ passing through the vertices $v_0$ and $v_k$ respectively.}
\label{mod}
\end{figure} 

\emph{Proof:}
Let $L_{0}$ be a line of symmetry for $G$ passing through the vertex $v_0$ of
$G$. Let the vertices starting from the vertex next
to $v_0$ in clockwise direction be $v_1, v_2, \ldots, v_m$. The angular
distances between the vertices of $G$, starting from $v_0$ in clockwise
direction, are denoted by 
${\theta}_0, {\theta}_1, \ldots, {\theta}_t, \ldots, {\theta}_{m}$ (Fig.
\ref{mod}). Since $L_{0}$ is 
a line of symmetry for $G$, ${\theta}_0 = {\theta}_{m}, {\theta}_1 =
{\theta}_{m-1}, \ldots,$ i.e.,
\begin{equation}
\label{1st}
 {\theta}_{i} = {\theta}_{(m - i){mod}(m)}   ~~~~~~~ 0 \le i \le \lfloor m/2
\rfloor
\end{equation}
 Suppose there is another line of symmetry passing through the vertex, $v_k$. 
This implies, ${\theta}_k = {\theta}_{k-1}, {\theta}_{k+1} = {\theta}_{k-2},
\ldots, {\theta}_{0} = {\theta}_{2k -1}, 
{\theta}_{1} = {\theta}_{2k-1}, \ldots$. We can represent the series by the
following equations: 
\begin{equation}
\label{2nd}
{\theta}_{j} = {\theta}_{(2k - j - 1){mod}(m)} ~~~~~~~ 0 \le j \le \lfloor m/2
\rfloor
\end{equation}
\begin{equation}
\label{3rd}
{\theta}_{m-j} = {\theta}_{(2k - j + 2){mod}(m)} ~~~~~~~ 0 \le j \le \lfloor m/2
\rfloor
\end{equation}

Let us consider any angle ${\theta}_i$ for $0 \le i \le m$. Combining 
equations \ref{2nd} and \ref{3rd} we get, ${\theta}_{i} = {\theta}_{(2k - i -
1){mod}(m)}$. Substituting ${\theta}_{(2k -i - 1){mod}(m)}$ in equation $3$ we
get ${\theta}_{(m - (2k -i-1)){mod}(m)} = 
{\theta}_{(2k - (2k - i-1)+2){mod}(m)} = {\theta}_{(i-1){mod}(m)}$. Therefore,
${\theta}_i = {\theta}_{i+1}$ for 
$0 \le i \le m-1$. This implies that, $G$ is regular.  
\qed

\begin{figure}[!h]

\centering
\includegraphics[ width=125mm]{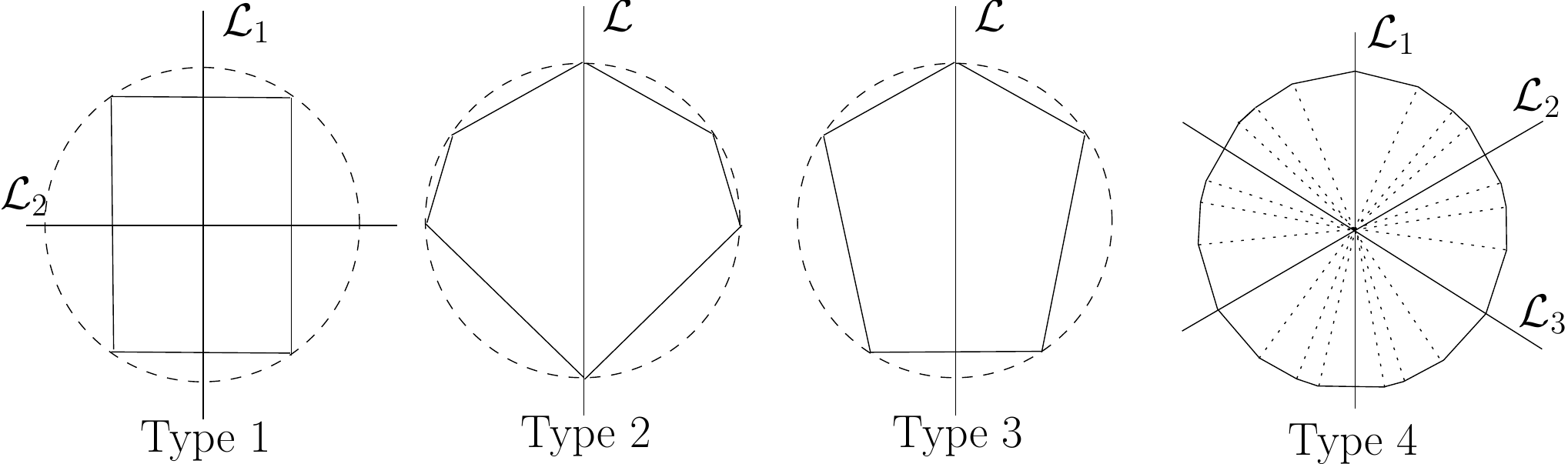}
\caption{Non regular symmetric polygon} 
\label{type3}
\end{figure}

Let us define different types of polygon on the basis of symmetry as follows.
\begin{itemize}
\item A {\bf type 0} polygon is a regular convex symmetric polygon. 

\item A {\bf type 1} polygon (Fig. \ref{type3}) is a convex, symmetric, non-regular 
polygon with even number of vertices and has exactly two
lines of straight symmetry $L_1$ and $L_2$, such that
\begin{itemize}
\item $L_i$ ($1 \le i \le 2$) does not pass through a vertex of the polygon.
\item $L_1 \bot L_2$.
\end{itemize}

It does not have any other line of straight symmetry but admits lines of skew symmetry.
  
\item  A {\bf type 2} polygon (Fig. \ref{type3}) is a convex, symmetric, non-regular 
polygon with even number of vertices and has exactly one line of
straight symmetry passing through either two vertices or two edges.

\item A {\bf type 3} polygon (Fig. \ref{type3}) is a convex, symmetric, non-regular 
polygon with odd number of vertices and has exactly one line of
straight symmetry passing through a vertex and an edge.

\item A {\bf type 4} polygon (Fig. \ref{type3}) is a convex, symmetric, non-regular 
polygon with odd number (not prime) of vertices such that the number of lines of symmetry is more than one but less than the number of vertices. 

Note : The number of lines of symmetry for this polygon is odd. 
\end{itemize}
 
\begin{obs}
\label{obs-exh-poly}
The above characterization of straight symmetric polygons is exhaustive. 
\end{obs}

\begin{obs}
\label{obs-sym-type-0}
If $G$ is a type 0 polygon with even number of vertices then any straight
line passing through the center of $G$ is a line of symmetry for $G$. 
\end{obs}

\begin{obs}
\label{obs-sym-type-01}
If $G$ is a type 0 polygon with odd number of vertices then any line of
symmetry must pass through exactly one vertex of $G$. 
\end{obs}

\begin{lemma}
\label{lem-sym-type-01}
Any straight line passing through the center of a type 1 polygon, is a line of symmetry (skewed or straight)
for that polygon.  
\end{lemma}

\emph{Proof:}
Let $G$ be a type 1 polygon. $G$ has exactly two lines of straight symmetry say ${\cal L}_1$ and ${\cal L}_2$. ${\cal L}_1$ and ${\cal L}_2$ pass through the center $C$ of $G$. It is sufficient to prove that any line between ${\cal L}_1$ and ${\cal L}_2$ is a line of skew symmetry for $G$.
Without loss of generality let us rotate ${\cal L}_1$ by some angle, around $C$, keeping it between ${\cal L}_1$ and ${\cal L}_2$. Let ${\cal L}'_1$ be the new position of the line. ${\cal L}'_1$ intersects $G$ either at two edges or passes through two vertices. 
If ${\cal L}'_1$ passes through two edges of $G$, then using lemma \ref{LoSkS-edge} and 
using an argument similar to that used in the proof of lemma \ref{lem-rectangle}, it can be shown that, polygonal chains on both sides of ${\cal L}'_1$ 
are skew symmetric. 
If ${\cal L}'_1$ passes through two vertices of $G$, then using lemma \ref{LoSkS-vertex} and 
using an argument similar to that used in the proof of lemma \ref{lem-rectangle}, it can be shown that, polygonal chains on both sides of ${\cal L}'_1$ 
are skew symmetric. 
Hence, ${\cal L}'_1$ is a line of skew symmetry.  
\qed

\begin{obs}
\label{obs-gcd}
Let $G_i$ and $G_j$ be two polygons with $n_i$ and $n_j$ vertices respectively. $n_i$ and $n_j$ are odd numbers.
If $<G_i$, $G_j>$ has one common line of symmetry, then $<G_i$, $G_j>$ has $gcd(n_i,n_j)$ many common lines of symmetry. 
\end{obs}

Let $G_i$ and $G_j$ ($i < j$) be two polygons in $\cal G$ with number of vertices $n_i$ and $n_j$ respectively such that 

\begin{itemize}
\item $n_i$, $n_j$ are odd.
\item $n_i$ does not divide $n_j$ or $n_j$ does not divide $n_i$   .
\item $gcd(n_i,n_j) > 1$.
\end{itemize}
$<G_i, G_j>$ has at least one common line of symmetry. We construct $G_{ij}$, and replace both $G_{i}$ and $G_j$ by $G_{ij}$ in $\cal G$. Following observation \ref{obs-gcd}, $G_{ij}$ has $gcd(n_i,n_j)$ many common lines of symmetry. 

\begin{lemma}
\label{lem-los-gij}
A line of symmetry for $G_{ij}$ is a common line of symmetry for $<G_i, G_j>$.
\end{lemma}

\emph{Proof:} The number of lines of symmetry of $G_{ij}$ is $gcd(n_i,n_j)$, where $n_i$ and $n_j$ are the numbers of vertices of $G_i$ and $G_j$ respectively. The angle between two adjacent lines of symmetry for $G_{ij}$ is always $\frac{360}{gcd(n_i,n_j)}$ degrees. The angle between two adjacent lines of symmetry of $G_i$ is $\frac{360}{n_i}$ degrees, which divides $\frac{360}{gcd(n_i,n_j)}$ degrees. This implies that these lines of symmetry of $G_{ij}$ are also the lines of symmetry for $G_i$. Using similar argument it can be stated that the lines of symmetry of $G_{ij}$ are also the lines of symmetry for $G_j$. Hence the result follows.
\qed

\begin{theorem}
\label{theorem-los-g1n}
A line of symmetry for $G_{1 \ldots n}$ is a common line of symmetry for $G_1, \ldots, G_n$. 
\end{theorem}

\emph{Proof:}
The statement is true for two concentric polygons (lemma \ref{lem-los-gij}). Suppose the result is true for $p$ polygons. Now ${p+1}^{th}$ polygon $G_{p+1}$ is introduced. 
If $G_{p+1}$ has even number of vertices then any common line of symmetry for $G_1, G_2, \ldots, G_p$, which will pass through the  center of the polygons, is also a line of symmetry for $G_{p+1}$. Hence the result is true.

Suppose $G_{p+1}$ has odd number of vertices.  Let ${\cal L}_1$ be a line of symmetry for $G_{1,2}$. From lemma \ref{lem-los-gij}, ${\cal L}_1$ is a common line of symmetry for $G_1$ and $G_2$. We merge $G_{1,2}$ and $G_3$ to get $G_{1,2,3}$. Let ${\cal L}_2$ be a line of symmetry for $G_{1,2,3}$. From lemma \ref{lem-los-gij}, ${\cal L}_2$ is a common line of symmetry for $G_{1,2}$ and $G_{3}$. Again, ${\cal L}_2$ is a common line of symmetry for $G_{1}$ and $G_{2}$. Proceeding further in this manner it can be shown that if ${\cal L}_{(n-1)}$ is a line of symmetry for $G_{1,\ldots,n}$ then ${\cal L}_{(n-1)}$ is a common line of symmetry for $G_1, \ldots, G_n$.
\qed

\begin{obs}
\label{obs-type 4}
If $n_i$ and $n_j$ are odd, $<G_i, G_j>$ has a common line of symmetry. Additionally, if $gcd(n_i,n_j)>1$, then $G_{ij}$ is a type 4 polygon.
\end{obs}

We repeat this process of merging polygons until every symmetric polygon with odd number of vertices becomes either of type 0 or type 3 or type 4. We get a modified version of $\cal G$, and call it ${\cal G}'$. 

\begin{lemma}
\label{lem-type4}
Let  $G_i$, $G_j$ and $G_k$ be three polygons in ${\cal G}'$ having odd number of vertices $n_i$, $n_j$ and $n_k$ respectively. No two of $n_i$, $n_j$ and $n_k$ are equal
and no one of $n_i$, $n_j$ and $n_k$ is a multiple of other. 
If $G_i$, $G_j$ and $G_k$ are pairwise symmetric then $n_j$, $n_j$ and $n_k$ are  prime to each other.
\end{lemma}

\emph{Proof:}
Suppose $n_j$, $n_j$ and $n_k$ are not prime to each other. Without loss of generality, let $gcd(n_i,n_j) > 1$. Hence, $G_i$ and $G_j$ can be merged to form $G_{ij}$, which is a type 4 polygon. This contradicts the construction process of $\cal G'$. 
\qed
\begin{lemma}
\label{lem-3-pair-sym}
Any three polygons $G_i$, $G_j$, and $G_k$ in ${\cal G}'$ are pairwise
symmetric if and only if $G_i$ , $G_j$ and $G_k$ have a common 
line of symmetry. 
\end{lemma}
\emph{Proof:}

{\bf If:} Trivial.

{\bf Only if:}
Since every pair is symmetric each individual polygon of $G_i$, $G_j$ and $G_k$ must be symmetric. Each polygon is either skew symmetric or type 0/1/2/3/4 polygon. 
Let ${\cal L}_{ij}$, ${\cal L}_{jk}$ and ${\cal L}_{ki}$ be three common lines of symmetry for $<G_i, G_j>$, $<G_j, G_k>$ and $<G_k, G_i>$ respectively. ${\cal L}_{ij}$, ${\cal L}_{jk}$ and ${\cal L}_{ki}$ pass through the common center ($C$) of $G_i$, $G_j$ and $G_k$.  
\begin{itemize}
\item If any of the polygons $G_i$, $G_j$ and $G_k$ (without loss of generality let it be $G_i$) is a skew symmetric polygon, then any line passing through $C$ is a line of symmetry for $G_i$ (lemma \ref{lem-skew-los}). As ${\cal L}_{jk}$ passes through $C$, it is also a line of symmetry for $G_i$. Therefore, ${\cal L}_{jk}$ is a common line of symmetry for $G_i$, $G_j$ and $G_k$. 

\item If any of the polygons $G_i$, $G_j$ and $G_k$ (without loss of generality let it be $G_i$) is a type 1 polygon, then any line passing through $C$ is a line of symmetry for $G_i$ (lemma \ref{lem-sym-type-01}). So ${\cal L}_{jk}$ is also a line of symmetry for $G_i$.
  
\item If any of the polygons $G_i$, $G_j$ and $G_k$ (without loss of generality let it be $G_i$) is a type 2 or type 3 polygon, then $G_i$ has exactly one line of symmetry. Hence, ${\cal L}_{ij} = {\cal L}_{ki}$. So the result follows.

\item If any of the polygons $G_i$, $G_j$ and $G_k$ (without loss of generality let it be $G_i$) is a type 0 polygon with even number of vertices, then any line passing through $C$ is a line of symmetry for $G_i$ (observation \ref{obs-sym-type-0}). So ${\cal L}_{jk}$ is also a line of symmetry for $G_i$.

\item If each of $G_i$, $G_j$ and $G_k$ is a type 0 polygon with odd number of vertices or type 4 polygon then following sub cases are possible:
 
\begin{itemize}
\item If any two polygons (without loss of generality, let them be $G_i$ and $G_j$) have equal number of vertices,
then all lines of symmetry of $G_j$ are also the lines of symmetry for $G_i$ and vice-versa.
Hence, ${\cal L}_{jk}$ is a line of symmetry for $G_i$. 

\item Suppose, the number of vertices of one polygon (say $G_i$) is a multiple of the number of vertices of another polygon (say $G_j$). Since, $G_i$ and $G_j$ share a common line of symmetry, every line of symmetry for $G_j$ is also a line of symmetry for $G_i$. So ${\cal L}_{jk}$ is a common line of symmetry for all three.    

\begin{figure}[!h]
\centering
\includegraphics[height=65mm, width=65mm]{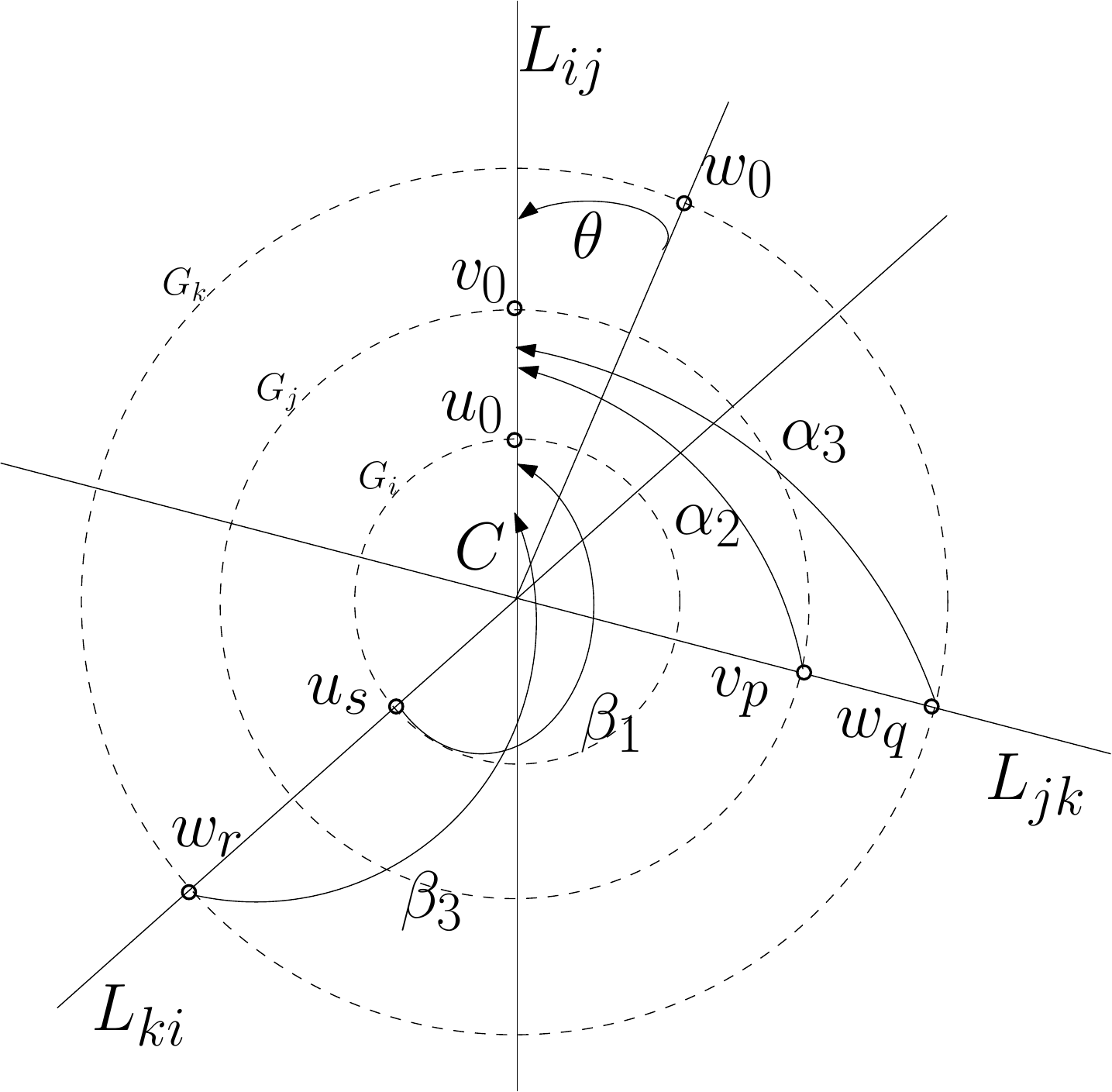}
\caption{Concentric polygon of odd number of vertices} 
\label{oddprime}
\end{figure}

\item Now we consider the case when none of the above is true. Let $n_i$, $n_j$ and $n_k$ be the number of vertices of $G_i$, $G_j$ and $G_k$ respectively (Figure \ref{oddprime}). $n_i$, $n_j$ and $n_k$ are prime to each other (lemma \ref{lem-type4}). 
${\cal L}_{ij}$ passes through a vertex $u_0$ of $G_i$ and a vertex $v_0$ of $G_j$. First we consider the case when both $u_0$ and $v_0$ lie on the same ray ($R_{ij}$) starting from $C$.    
Suppose, ${\cal L}_{ij}$ does not pass through any vertex of $G_k$. Let $w_0$ be the vertex of $G_k$ which is closest to ${R}_{ij}$ in the clockwise direction.
$Cw_0$ makes an angle $\theta$ with $R_{ij}$ at $C$. We label the vertices of $G_i$ in the clockwise direction starting from $u_0$ and denote them by $u_1,\ldots,u_{n_{i-1}}$. We label the vertices of $G_j$ in the clockwise direction starting from $v_0$ and denote them by $v_1,\ldots,v_{n_{j-1}}$. We label the vertices of $G_k$ in the clockwise direction starting from $w_0$ and denote them by $w_1,\ldots,w_{n_{k-1}}$.     
Suppose, ${\cal L}_{jk}$ passes through the $p^{th}$ vertex $v_p$ of $G_j$ and the $q^{th}$ vertex $w_q$ of $G_k$. 
${\cal L}_{ki}$ passes through the vertex $w_r$ of $G_k$ and $u_s$ of $G_i$. 
$Cv_p$ and $Cw_q$ make angles $\alpha_2$ and $\alpha_3$ respectively with $R_{ij}$. 
$Cu_s$ and $Cw_r$ make angles $\beta_1$ and $\beta_3$ respectively with $R_{ij}$.  

Since, $\alpha_2 = \alpha_3$, we get the following equation:
\begin{equation}
\frac{p}{n_j} = \frac{q}{n_k} + \frac{\theta}{360}
\end{equation}
Since, $\beta_1 = \beta_3$, 
\begin{equation}
\frac{s}{n_i} = \frac{r}{n_k} + \frac{\theta}{360} 
\end{equation}

Substituting equation $(4)$ from equation $(5)$,
 
\begin{equation}
({\frac{s}{n_i} - \frac{p}{n_j}}){n_k} = r - q
\end{equation}

Since $(r-q)$ is an integer, $ ({\frac{s}{n_i} - \frac{p}{n_j}}){n_k}$ must be an integer. 
$0 \le \frac{s}{n_i} < 1$ and $0 \le \frac{p}{n_j} < 1$. $|({\frac{s}{n_i} - \frac{p}{n_j}})| < 1$.
Since, $n_i$, $n_j$ and $n_k$ are pairwise relatively prime, $({\frac{s}{n_i} - \frac{p}{n_j}}){n_k} = 0$.   
From equation $(4)$, $\frac{s}{n_i} = \frac{p}{n_j} = \frac{q}{n_k} + \frac{\theta}{360}$. 
Hence, ${\cal L}_{jk}$ is the line of symmetry for $G_i$, $G_j$ and $G_k$. 
\end{itemize}
\end{itemize}

\begin{figure}[!h]
\centering
\includegraphics[height=65mm, width=65mm]{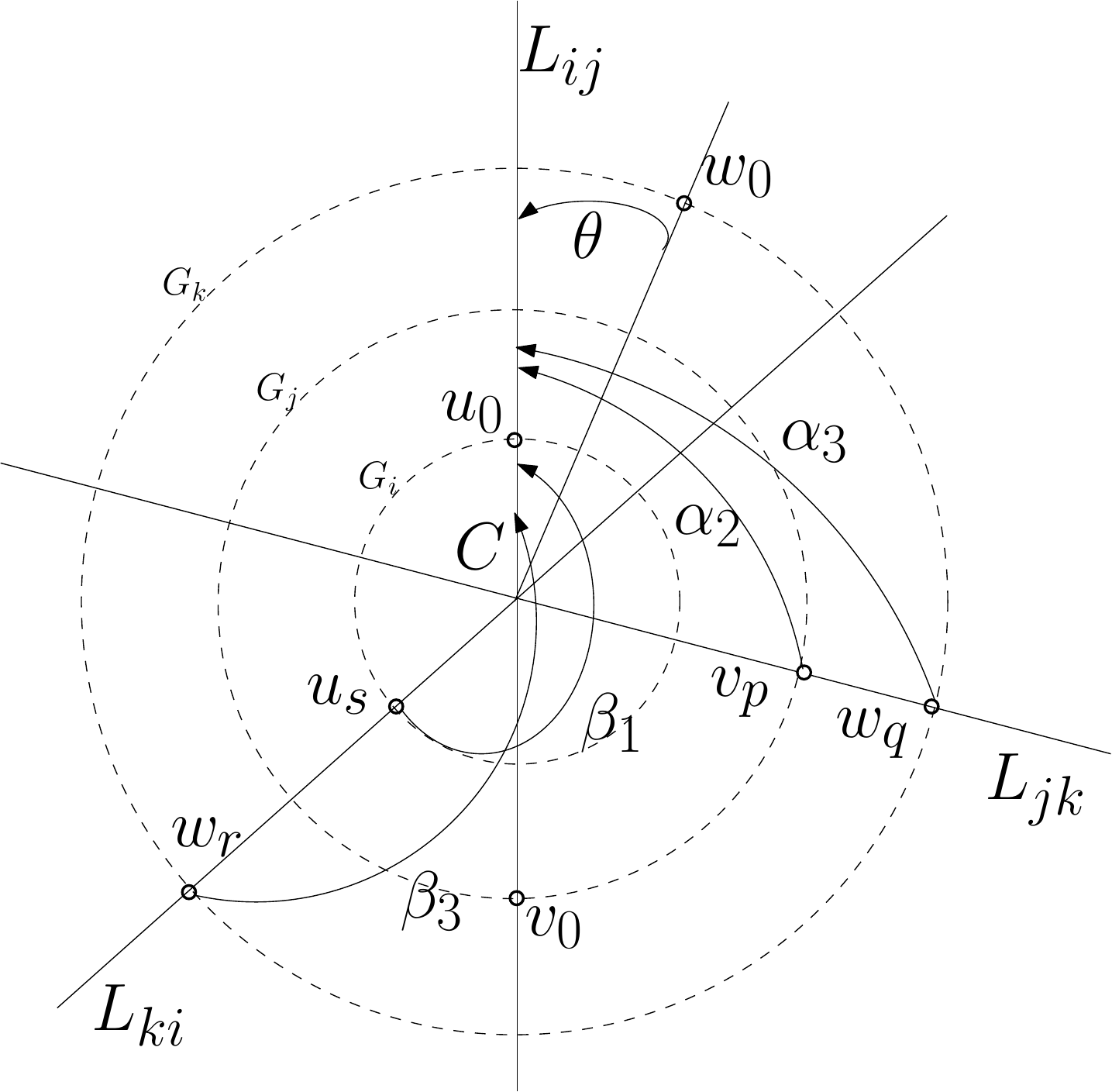}
\caption{Concentric polygon of odd number of vertices} 
\label{oddprime1}
\end{figure}

Let us consider the case when $u_0$ and $v_0$ lie at different sides of $C$ (Fig \ref{oddprime1}). The equations $(4)$ and $(5)$ will now change.
 
Since, $\alpha_2 = \alpha_3$, we get the following equation:
\begin{equation}
\frac{p}{n_j} = \frac{q}{n_k} + \frac{\theta}{360} + \frac{180}{360}
\end{equation}
Since, $\beta_1 = \beta_3$, 
\begin{equation}
\frac{s}{n_i} = \frac{r}{n_k} + \frac{\theta}{360} 
\end{equation}

Substituting equation $(7)$ from equation $(8)$,
 
\begin{equation}
({\frac{s}{n_i} - \frac{p}{n_j}} + \frac {1}{2}){n_k} = r - q
\end{equation}

From the argument stated previously, the value of ${\frac{s}{n_i} - \frac{p}{n_j}}$ is a fraction. 
If  ${\frac{s}{n_i} - \frac{p}{n_j}} \ne 0$, then it has a fractional part. 
So, $|({\frac{s}{n_i} - \frac{p}{n_j}} + \frac {1}{2}){n_k})|$ is not a non-zero integer. Since, $r-q$ is an integer, it must be zero.  
From equation $(7)$, $\frac{s}{n_i} = \frac{p}{n_j} = \frac{q}{n_k} + \frac{\theta}{360} + \frac{1}{2}$. 
Hence, ${\cal L}_{jk}$ is a common line of symmetry for $G_i$, $G_j$ and $G_k$. 

Similarly the result follows for other cases such as when $v_p$ and $w_q$ lie on different sides of $C$ or $u_s$ and $w_r$ lie on different sides of $C$.    
\qed

\begin{theorem}
\label{theorem-sym-pair-symG}
Every pair in $\cal G'$ is a symmetric pair, if and only if there is a common line of symmetry for all the polygons
in $\cal G'$.
\end{theorem}
\emph{Proof:}
{\bf If:} Trivial.

{\bf Only If:} We prove this by induction on the number of polygons in $\cal G'$. 
If $\cal G'$ contains three polygons then the result follows from lemma \ref{lem-3-pair-sym}. 
Suppose, the statement is true for $p \ge 3$ polygons. Now the ${p+1}^{th}$ polygon say $G_{p+1}$ is introduced. If $G_{p+1}$ has even number of vertices then any line of symmetry for $G_{1\ldots p}$ is a line of symmetry for $G_{p+1}$. Hence the result is true.

Suppose $G_{p+1}$ has odd number of vertices. 
$<G_p, G_{(p+1)}>$ is a symmetric pair (given). Since $G_1, G_2, \cdots, G_p$ has a common line of symmetry (induction hypothesis), that same line is also a common line of symmetry for $<G_{1,2,\cdots,(p-1)}, G_p>$. Similarly, using the induction hypothesis $<G_{(p+1)}, G_{1,2,\cdots,(p-1)}>$ is also a symmetric pair. Hence following lemma \ref{lem-3-pair-sym}, $G_{1,2,\cdots,(p-1)}$, $G_p$ and $G_{(p+1)}$ have a common line of symmetry, say $\cal L$. Following theorem \ref{theorem-los-g1n}, $\cal L$ is a line of symmetry for $G_1, \ldots G_{(p-1)}$. Thus the result follows.  
\qed      

 \begin{obs}
 \label{obs1-G'equiG}
 If $\cal G$ is symmetric then every pair in $\cal G$ is a symmetric pair.
 \end{obs}

\begin{theorem}
\label{G'equiG}
$\cal G'$ is symmetric if and only if $\cal G$ is symmetric.
\end{theorem}

\emph{Proof:} Follows from theorem \ref{theorem-los-g1n}.
\qed

Let $P$ be a set of points on the 2D plane. We first identify the SEC for $P$. With respect to the center of the SEC we divide the points of $P$ into a set of concentric polygons ${\cal G}_P$. 

\begin{theorem}
\label{Psym-Gsym}
$P$ is in $\O_{S}$ iff ${\cal G}_P$ is in $\O_{S}$. 
\end{theorem}

\emph{Proof:}{\bf If:} If ${\cal G}_P$ is in $\O_{S}$, all polygons 
in $\cal G$ have a common line of symmetry (definition \ref{def-sym-calG}).
Hence, the vertices in $G_1 \cup G_2 \ldots \cup G_t$ have a line of
symmetry. The set of vertices in $G_1 \cup G_2 \ldots \cup G_t$, which is $P$, is in $\O_{S}$. 

{\bf Only If:} Let $\cal L$ be a line of symmetry for $P$. $\cal L$ also passes
through the center of SEC of $P$ (observation \ref{lemma-LOS-Passes-CenterSECP}).
However, $P$ is the set of vertices in $G_1 \cup G_2 \ldots \cup G_t$. $\cal L$
also is the line of symmetry of the set of vertices in $G_1 \cup G_2 \ldots \cup
G_t$. It is easy to note that the mirror image of any vertex of $G_i \in \cal
G$, is a vertex of ${\cal G}_i$. This implies that, $\cal L$ is the line of
symmetry for all $G_i \in \cal G$. Therefore all $G_i \in \cal
G$, are symmetric across $\cal L$. Hence, ${\cal G}_P$ is in $\O_{S}$ (definition
\ref{def-sym-calG}). 
\qed

\begin{corollary}
$P$ is in $\O_{A}$ iff ${\cal G}_P$ is in $\O_{A}$.
\end{corollary}

\begin{theorem}
\label{Pasym-Pord}
$P$ is orderable if and only if $P$ is in $\O_{A}$.  
\end{theorem}
\emph{Proof:} {\bf If:} If $P$ is in $\O_{A}$, ${\cal G}_P$ is in $\O_{A}$ (theorem
\ref{Psym-Gsym}). If ${\cal G}_P$ is in $\O_{A}$, there is no common line of symmetry for the polygons in ${\cal G}_P$ (definition \ref{def-sym-calG}). There exists an asymmetric pair in ${\cal G}_P$. 
Following lemma \ref{lem-all-pair-asym-ord}, ${\cal G}_p$ is orderable. Hence $P$ is orderable.

{\bf Only If:} If $P$ is orderable, $P$ is in $\O_{A}$ (lemma 1).
\qed

\begin{corollary}
\label{cor-Pasym-Pord}
$P$ is orderable if and only if ${\cal G}_P$ is in $\O_{A}$.
\end{corollary}

\section{Algorithms for Leader Election and Gathering of Robots}
\label{algo}
Let $R$ be a set of robots. From the previous section, we note that if
$R$ is orderable, then leader election is possible from $R$. In this case the
first robot in the ordering becomes the leader. In this section, we present the
leader election algorithm for a set of robots, $R$. $R$ can be viewed as a set
of multiple concentric polygons ${\cal G} = G_1, G_2, \ldots, G_t$ as described
previously. The leader election algorithm elects leader from $G_1 \in {\cal
G}$. First we present leader election algorithm for a single polygon $G$. Then
we extend the algorithm for the set of concentric polygons $\cal G$. 

\subsection{Algorithm for checking symmetry in $G$}
\label{checksym_algo_G}
A polygon may have more than one line of symmetry. 
\begin{defi}
The number of lines of symmetry of $G$ is called the degree of symmetry of $G$.
\end{defi}

Algorithms $Check\_Symmetry\_Odd(G)$ and $Check\_Symmetry\_Even(G)$ are used for checking symmetry in $G$ when the numbers of vertices of $G$ are odd and even respectively.

\begin{algorithm}[H]
\SetLine
\linesnumbered
\KwIn{ A convex polygon $G$ with $n$ vertices such that $n$ is odd. }
\KwOut{Reports the degree of symmetry of $G$ ($0$ if asymmetric)}
$deg\_sym = 0$\;
$i = 1$\;
\While{$i \le n$} {
Compute $CW_{v_i}$ and $ACW_{v_i}$\;
\eIf{$CW_{v_i} = ACW_{v_i}$}{
$deg\_sym = deg\_sym + 1$\;
}
{$i++$ \;}
}
\eIf{$deg\_sym > 0$}
{Report ``$G$ is symmetric''\;
 Report ``The degree of symmetry is:'' $deg\_sym$\;}
{Report ``$G$ is asymmetric''\;}
\caption{$Check\_Symmetry\_Odd(G)$}
\end{algorithm} 

{\it Correctness of Check\_Symmetry\_Odd(G):} Follows from theorem \ref{obs-asymmetric-polygon}.
\qed

\begin{algorithm}[H]
\SetLine
\linesnumbered
\KwIn{A convex polygon $G$ with $n$ vertices such that $n$ is even}
\KwOut{Reports the degree of symmetry of $G$ ($0$ if asymmetric)}

$deg\_sym = 0$\;
$i = 1$\;
$st\_sym = False$\;
$sk\_sym = False$\;

\While{$i \le n$} {
Compute $CW_{v_i}$ and $ACW_{v_i}$\;
\eIf{$CW_{v_i} = ACW_{v_i}$}{
$st\_sym = True$\;
$deg\_sym = deg\_sym + 1$\;
}
{\For{j = 1 to i-1}
 {\If{($CW_{v_i} = CW_{v_j}$) $\vee$  ($CW_{v_i} = ACW_{v_j}$) $\vee$ ($ACW_{v_i} = CW_{v_j}$) }{
$sk\_sym = True$\;
$deg\_sym = deg\_sym + 1$\;
}
}
$i++$\;
}
}
\If{$deg\_sym > 0$ $\wedge$ $st\_sym = True$}
 {Report ``$G$ is straight symmetric''\;
 Report ``The degree of symmetry is:'' $deg\_sym/2$\;}
\If{$deg\_sym > 0$ $\wedge$ $sk\_sym = True$}
 {Report ``$G$ is skew symmetric''\;
 Report ``The degree of symmetry is:'' $deg\_sym/2$\;}
\If{$deg\_sym = 0$}
{Report ``$G$ is asymmetric''\;}
\caption{$Check\_Symmetry\_Even(G)$}
\end{algorithm} 

{\it Correctness of Check\_Symmetry\_Even(G):} Follows from theorem \ref{obs-asymmetric-polygon}.
\qed

Note that if $G$ has even number of vertices then every line of symmetry (straight or skewed) passes through 2 vertices and is counted twice. Therefore, the algorithm returns the $deg\_sum$ by dividing it by $2$.

Next we present a leader election algorithm, when the convex polygon has degree of symmetry one. 

\begin{algorithm}[H]
\SetLine
\linesnumbered
\KwIn{ A symmetric convex polygon $G$ of $n$ vertices.}
\KwOut{A vertex $v_l$ or reports that leader election is not possible.}
\eIf{$n$ is odd}
{Call $Check\_Symmetry\_Odd(G)$\;
\If{degree of symmetry of $G > 1$}
{Report ``Leader Election is not possible for $G$''\; }
\If{degree of symmetry of $G$ = $1$}
{$v_l$ =  $v_i \in G$ such that $CW(v_i) = ACW(v_i)$\;
Return $v_l$\;}
}
{
Call $Chekc\_Symmetry\_Even(G)$\;
\If{degree of symmetry of $G > 1$}
{Report ``Leader Election is not possible for $G$''\; }
\eIf{degree of symmetry of $G$ = $1$ and the line of symmetry passes through either a single vertex or through two vertices.}
{Find two vertices $v_i$ and $v_j$ such that $CW(v_i) = ACW(v_i)$ and $CW(v_j) = ACW(v_j)$\;
$VI = CW(v_i)$\;
 $VJ = CW(v_j)$\;
\eIf{$VI > VJ$}
{return $v_i$ as leader\;}
{return $v_j$ as leader\;}
}
{Report ``Leader Election is not possible for $G$''\;}
}
\caption{$Elect\_Leader\_Sym(G)$}
\end{algorithm} 

{\it Correctness of Elect\_Leader\_Sym(G):} Follows from  theorem \ref{obs-asymmetric-polygon}.
\qed

\begin{obs}
 \label{1sym_le}
If the degree of symmetry of a convex symmetric polygon is one and the line of symmetry passes through two vertices of the polygon, then leader election is possible.
\end{obs}

Once the leader is elected for a symmetric polygon with degree of symmetry one, the leader can be moved such a way 
that the new polygon becomes asymmetric.
Following algorithm does this task.

\begin{algorithm}[H]
\scriptsize
\SetLine
\linesnumbered
\KwIn{ A convex polygon $G$ with degree of symmetry one. The line of symmetry passes through either a single vertex or through two vertices.}
\KwOut{An asymmetric polygon convex $G$.}
$v_l$ = $Elect\_Leader\_Sym(G)$\;
Move $v_l$, $\epsilon$ distance, to its right side, on the circumference of the circle inscribing $G$\;
return $G$\;
\caption{$Make\_SymToAsym(G)$}
\end{algorithm}

{\it Correctness of Make\_SymToAsym(G):} Whenever the leader $v_l$ moves from its position, $G$ becomes asymmetric. Then no other 
robot executes the algorithm $Make\_SymToAsym(G)$. Therefore, there is no chance that $G$ becomes symmetric again. 
\qed

Following algorithm elects leader when $G$ is asymmetric.

\begin{algorithm}[H]
\SetLine
\linesnumbered
\KwIn{ An asymmetric convex polygon $G$ with $n$ vertices.}
\KwOut{A leader vertex $v_l$. The new positions of $G \setminus v_l$ are asymmetric.}
\For{k =1 to n}
{Compute the tuple $\{CW(v_k), ACW(v_k)\}$\;}
$\{CW(v_m), ACW(v_m)\}$ = lexicographic minimum of $\{ (CW(v_1), ACW(v_1)), \ldots, (CW(v_n), ACW(v_n))\}$\; 
\eIf{$CW(v_m) < ACW(v_m)$}
{$CW(v_m)$ is the ordering of $G$\;}
{$ACW(v_m)$ is the ordering of $G$\;}
$i = 1$\; 
\While{$G \setminus v_i$ is symmetric}{
$i++$\;
}
$v_l$ =  $v_i$ \;

return $v_l$\;
\caption{$Elect\_Leader\_Asym(G)$}
\end{algorithm} 

{\it Correctness of Elect\_Leader\_Asym(G):} Follows from (lemma \ref{lem-asym-ord}).  
\qed

\subsubsection{ Algorithm for electing leader in $\cal G$}
\label{lederelection in _calG}
Let $\cal G$ be a set of $t$ concentric polygons as described previously. $G_1$ is the 
nearest to the center. 
$G_t$ is the farthest from the center. Algorithm $ElectLeader(\cal G)$ selects a leader from $\cal G$.  
The algorithm assumes that $G_t$ is asymmetric. Therefore, $<G_1, G_t>$ is an
asymmetric pair (Observation \ref{obs-pair-asym-ord}). 
$ElectLeader(\cal G)$ checks if $G_1$ is asymmetric. If $G_1$ is asymmetric, then the leader 
is selected from $G_1$ by algorithm $ElectLeader(G_1)$. If $G_1$ is symmetric, then $ElectLeader(\cal G)$ computes $G_{1t}$. Since, $G_t$ is asymmetric, 
$G_{1t}$ is orderable. Leader election is possible from $G_{1t}$.  

\begin{algorithm}[H]
\SetLine
\linesnumbered
\KwIn{$\cal G$ such that $G_t$ is asymmetric.}
\KwOut{A leader vertex $v_l$ from $G_1 \in \cal G$ such that $G_1 \setminus v_l$ is asymmetric.}
\eIf{the number of vertices in $G_1$ is even}
{Call $Chekc\_Symmetry\_Even(G_1)$\;}
{Call $Chekc\_Symmetry\_Odd(G_1)$\;}
\eIf{$G_1$ is symmetric with degree of symmetry one and the line of symmetry passes through either a single vertex or through two vertices.}
{Call $Make\_SymToAsym(G_1)$\;
$v_l = Elect\_Leader\_Asym(G_1)$\;
Report ``the leader $v_l$''\; 
}
{\eIf{$G_1$ is asymmetric}
{$v_l = Elect\_Leader\_Asym(G_1)$\;
Report ``the leader $v_l$''\; 
}
{Compute $G_{1t}$\;
$v_l = Elect\_Leader\_Asym(G_{1t})$\;
Report ``the leader $v_l$''\; 
}
}
\caption{$ElectLeader(\cal G)$ } 
\end{algorithm} 

{\it Correctness of ElectLeader($\cal G$):} 
Follows from lemma \ref{lem-one-pair-asym-ord}. 
\qed

\subsubsection{Algorithm for preprocessing $\cal G$ to make it asymmetric}
\label{preprocessing_calG}
First $\cal G$ is reconstructed to form $\cal G'$. 

\begin{algorithm}[H]
\scriptsize
\SetLine
\linesnumbered
\KwIn{$\cal G$.}
\KwOut{$\cal G'$.}
$i=1$\;
\While{$i \le t$}{
\eIf{$G_i$ has odd number of vertices}{
$j=1$\;
\While{$j \le t$}{
\eIf{$j \ne i$ $\&$ $G_j$ has odd number of vertices}
{Compute polygon $G_{ij}$\;
\eIf{$G_{ij}$ is a type 4 polygon}
{
Replace $G_{i}$ by $G_{ij}$\;
}
{$j++$;
}
}
{$j++$;
}
}
}
{
$i++$;
}
}
\caption{$Preprocess(\cal G)$ }  
\end{algorithm}

First $\cal G$ is preprocessed to generate $\cal G'$. Algorithm $ElectLeader(\cal G')$ selects a leader from $\cal G'$, assuming that $G_t$ in $\cal G'$ is asymmetric. 
The algorithm $MakeAsym(G_t)$ pre-processes $\cal G'$ to make $G_t$ asymmetric, whenever that is possible. 
Let $G_t$ in $\cal G'$ be symmetric. If any $G_i$ in $\cal G'$ is asymmetric, then $\cal G'$ is asymmetric (lemma \ref{lem-one-pair-asym-ord}). 
If there is no asymmetric polygon in $\cal G'$, but their exists a $G_i$ such that $G_i$ contains only one vertex $v$, and the vertex is on the line of symmetry of $G_t$, then $v$ is moved slightly ($\epsilon$ distance) form its current position. $<G_i, G_t>$ becomes asymmetric. 
$G_{it}$ is computed. The first vertex from $G_t$, in the ordering of $G_{it}$ is selected as a leader from $G_t$. The leader is 
moved to a new position ($\epsilon$ distance away from the current position) to make $G_t$ asymmetric. 
$Find\_Asymmetry(\cal G')$ finds an asymmetric polygon $G_a$ from $\cal G'$ using binary search on the set $G_1 \cup G_2 \ldots \cup G_t$. 
If $Find\_Asymmetry(\cal G')$ can not find any asymmetric polygon $\cal G'$ then an asymmetric pair $<G_i,G_j>$ is searched using algorithm  
$Find\_Asymmetry\_Pair(\cal G')$.
If any $<G_i, G_j>$ in $\cal G'$ is asymmetric, then $\cal G'$ is asymmetric (lemma \ref{lem-all-pair-asym-ord}). In such a case, 
$G_{ijt}$ is computed by radially projecting the polygon $G_{ij}$ on $G_t$. The first vertex from $G_t$, 
in the ordering of $G_{ijt}$ is selected as a leader from $G_t$. 
The leader is moved to a new position ($\epsilon$ distance away from the current position) to make $G_t$ asymmetric.
If no such pair is found, then $\cal G'$ is symmetric and hence not orderable.
The algorithm also maintains the number of vertices of $G_t$ to be more than $3$. This assures that the SEC will not 
change during the motion of the leader from $G_t$. 
Finally, $MakeAsym(G_t)$ adjusts $G_t$ to be asymmetric if it is initially symmetric.      

\begin{algorithm}[H]
\scriptsize
\SetLine
\linesnumbered
\KwIn{$\cal G'$, with $G_t$ is symmetric.}
\KwOut{An orderable $\cal G'$ where $G_t$ is asymmetric or report if $\cal G'$ is not orderable.}
$G_a = FindAsymmetry(\cal G')$\;
\eIf{asymmetric polygon $G_a$ is found}
{\eIf{$|G_t| > 3$}
{$v_l = Elect\_Leader\_Asym(G_{at})$ (such that $v_l$ is in $G_t$)\;
Move $v_l$, $\varepsilon$ distance away from the current position on the same circle\; 
Report ``An orderable $\cal G'$ where $G_t$ is asymmetric''\;
}
{Call CheckSymmetry($G_{t-1}$)\;
\eIf{$G_{t-1}$ is asymmetric}{
$v_l = Elect\_Leader(G_{t-1})$\;
}
{Call $ElectLeader( G_{a(t-1)})$\;}
Move $v_l$ to the circumference of the enclosing circle of $G_t$\;
Report ``An orderable $\cal G'$ where $G_t$ is asymmetric''\;
}
}
{
$<G_i,G_j> = Find\_Asymmetry\_Pair(\cal G')$\;
\eIf{asymmetric pair $<G_i,G_j>$ is found}
{\eIf{$|G_t| > 3$}
{$v_l = Elect\_Leader(G_{ijt})$\;
Move $v_l$, $\varepsilon$ distance away from the current position\; 
Report ``An orderable $\cal G$ where $G_t$ is asymmetric''\;
}
{Call CheckSymmetry($G_{t-1}$)\;
\eIf{$G_{t-1}$ is asymmetric}{
$v_l = Elect\_Leader(G_{t-1})$\;
}
{Call $ElectLeader(G_{ij(t-1)})$\;}
Move $v_l$ to the circumference of the enclosing circle of $G_t$\;
Report ``An orderable $\cal G'$ where $G_t$ is asymmetric''\;
}
}
{
\eIf{there exists a polygon $G_i \in {\cal G'}$ such that $G_i$ contains a single vertex $v$}
{Move $v$, $\varepsilon$ distance from its current position\;
$v_l = Elect\_Leader\_Asym(G_{it})$ (The first vertex from $G_t$ in the ordering of $G_{it}$)\;
Move $v_l$, $\varepsilon$ distance away from the current position\; 
}
{
Report ``$\cal G'$ is not orderable''\;
}
}

}
\caption{ $MakeAsym(\cal G)$} 
\end{algorithm}
\begin{obs}
 \label{obs-v-move}
Suppose there exists no asymmetric polygon and asymmetric pair, but there exists a polygon $G_i \in {\cal G'}$ such that $G_i$ contains a single vertex $v_l$. 
In this case, the leader $v_l$ selected from $G_i$ is same no matter whether $v_l$ is in motion or $\varepsilon$ distance apart from its position in symmetry.   
\end{obs}
\begin{lemma}
 \label{lem-vl-move}
When $v_l$ is moving no other robot will be selected as leader.
\end{lemma}
\emph{Proof:}
$v_l$ is selected either from $G_t$ or $G_{t-1}$.
If $v_l$ is selected from $G_t$, then after $v_l$ starts moving, $G_t$ becomes asymmetric. No robot will execute $MakeAsym(\cal G')$ again. Therefore $v_l$ remains the leader.
If $v_l$ is selected from $G_{t-1}$, then after $v_l$ starts moving, $v_l$ forms the new $G_{t-1}$ where $v_l$ is the only robot in $G_{t-1}$. Hence, $v_l$
remains the leader. 
\qed   

{\it Correctness of MakeAsym($G_t$):} The algorithm always maintains the 
number of vertices of $G_t$ to be greater than three. This ensures that the SEC 
would not change during the motion of any robots from $G_t$. Following lemma \ref{lem-all-pair-asym-ord}, lemma \ref{lem-one-pair-asym-ord} and observation \ref{obs-v-move}, if their is an asymmetric polygon or an asymmetric pair or a polygon with single vertex on the line of symmetry of $\cal G'$, then $\cal G'$ is orderable and leader election is possible.  Lemma \ref{lem-vl-move} assures that when the leader is moving, no other robots will move. 
\qed

\subsection{Algorithm for Gathering}
\label{gathering_algorithm}
In this section, the algorithm $GatheringPattern(\cal R)$ for gathering $n$ $( n \ge 5 )$ transparent
fat robots is presented. Initially all robots are assumed to be stationary and separated. Let $R$ be a set of 
such robots. Let $\cal P$ be the set of robots already in the gathering pattern.
Initially $\cal P$ is null. The algorithm first finds the center $C$ 
of the SEC for the robots in $R$. The robots in $R$ are ordered based on their distances from 
$C$. The robots equidistant from $C$ are on the circumference of a circle and form a 
convex polygon. The robots are the vertices of the polygons. Hence, we get a number of concentric polygons 
having a common center $C$. The robots equidistant from $C$, are said to be in the same level. 
Let there be $t$ levels of distances. The robots nearest 
to $C$ are at level $1$ and the robots farthest from $C$ are at level $t$. Let $\cal G$ 
be the set of these $t$ concentric polygons. $\cal G$ is preprocessed to generate $\cal G'$ using $Preprocess(\cal G)$. 

\begin{figure}[!h]
\centering
\includegraphics[height=45mm, width=120mm]{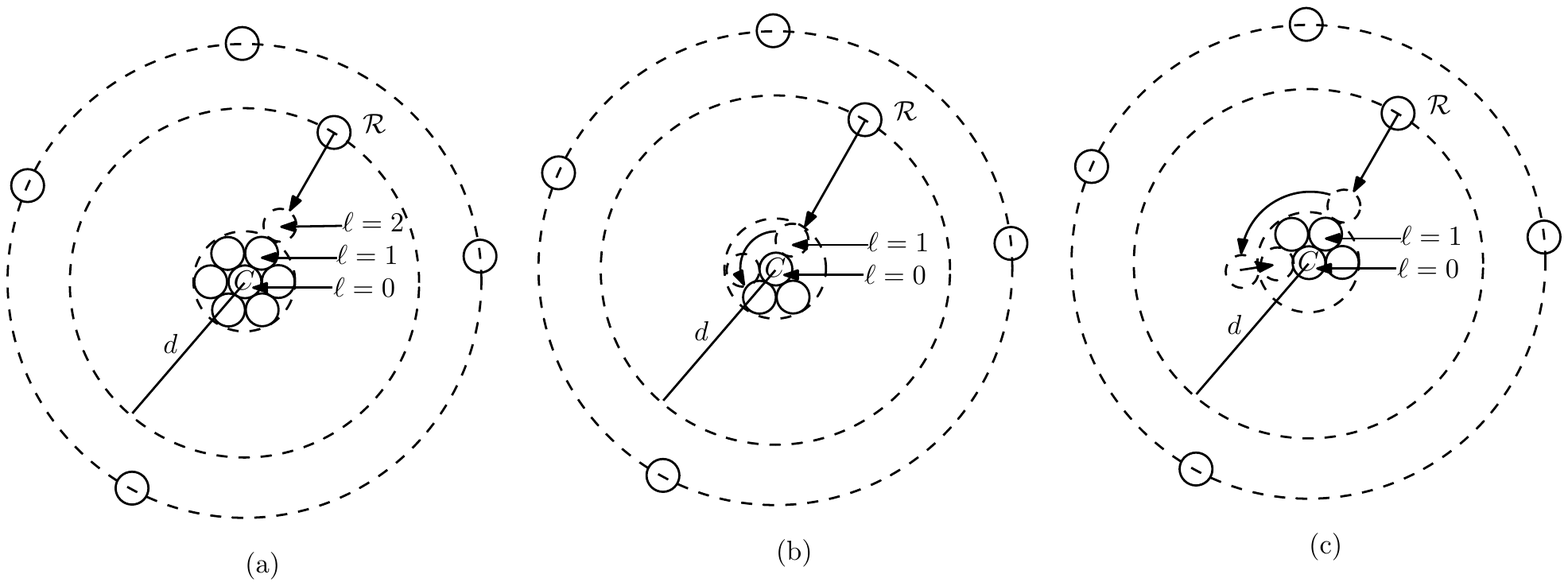}
\caption{Formation of the gathering pattern around the center $C$ of the
\emph{SEC}}
\label{gatheringalgoF}
\end{figure}

The gathering algorithm then processes $\cal G'$ using $MakeAsym(G_t)$ to make $G_t$ asymmetric. 
If $MakeAsym(G_t)$ reports that $\cal G'$ is in $\O{}_S$, then gathering is not possible. If $MakeAsym(G_t)$ 
returns an orderable $\cal G$ with $G_t$ asymmetric, then $GatheringPattern(\cal R)$
selects a leader robot $\cal R$ using $ElectLEader({\cal G})$ and 
moves $\cal R$ to extend the gathering pattern already formed, using algorithm $FormGPattern(\cal R)$. 
$\cal R$ moves towards $C$ in order 
to build the desired gathering pattern. If $C$ is not occupied by any other robot, then $\cal R$ becomes the central 
robot of the gathering pattern. If $C$ is occupied by another robot, then the algorithm finds the last layer $\ell$ of the 
gathering pattern. If $\ell$ is full, then $\cal R$ makes the new layer $\ell+1$
(Figure. \ref{gatheringalgoF}). If $\ell$ is not 
full, then $\cal R$ may slide around $C$ (if required) and is placed in layer $\ell$. When $\cal R$ reaches its destination and stops, 
the robot is removed from $R$ and added to $\cal P$. 

\begin{algorithm}[H]
\SetLine
\linesnumbered
\KwIn{A robot $\cal R$ }
\KwOut{ $\cal R$ with its final position in $\cal P$.}
\eIf {$C$ is not occupied by other robot}{
              Move $\cal R$ to $C$ \;}
{

$\ell$ = the last layer of $\cal P$ formed around $C$\;

            \eIf{$\ell$ is full}{
	     Move $\cal R$ to a new layer($\ell+1$)\;}
{
Move $\cal R$ to layer $\ell$\;
Slide $\cal R$ around $C$ (if required) in layer $\ell$ in order to touch other robot in layer $\ell$\; 
}
}
Remove $\cal R$  from $R$\; 
Add $\cal R$ to $\cal P$\; 
\caption{$FormGPattern(\cal R)$}
\end{algorithm}

\begin{algorithm}[H]
\SetLine
\linesnumbered
\KwIn{$R$, the set of separated robots.}
\KwOut{ $\cal P$, the gathering pattern.}
Calculate $SEC$ by the robots in set $R$. 
$C$ be the center of $SEC$\;
Build the set $\cal G$ with the robots in set $R$\;
$\cal G'$ = $Preprocess(\cal G)$\; 
\eIf{$\cal G'$ is in $\O{}_S$}
{Report that gathering is not possible\;
}
{ Call $MakeAsym(G_t)$\;
$r_l$ = $ElectLeader(\cal G')$\;
Call $FormGPattern(r_l)$\;
}
\caption{$GatheringPattern(\cal R)$}
\end{algorithm}

The following $two$ lemmas shows that the choice of the destination and the
robot for movement remain invariant during the motion of the designated robot.
\begin{figure}[!h]
\centering
\includegraphics[height=50mm, width=50mm]{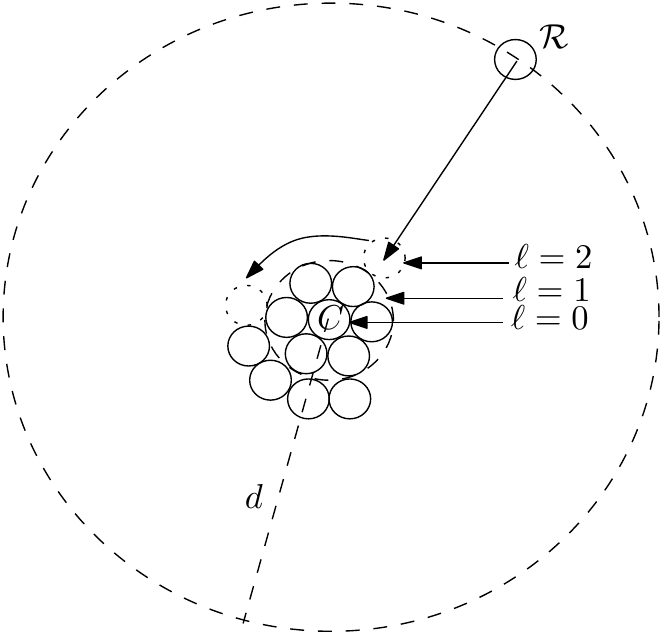}
\caption{The distance of a mobile robot from $C$ remains minimum during its
movement}
\label{invariant}
\end{figure}
\begin{lemma}
When a robot is moving towards $C$ (the center of SEC) and/or  sliding
around $C$, its distance from $C$ is still
minimum among all the robots in set $R$.
\end{lemma}
\emph{Proof}:
Let $\cal R$ be a robot, which is selected for moving towards $C$. Let $d= \overline{\cal R,C}$ (Figure. \ref{invariant}). No other robot is inside the circle of radius $d$.
According to $GatheringPattern(\cal R)$, only $\cal R$
starts moving towards $C$ and other robots are stationary. $\cal R$ moves
towards $C$ and after touching the
robots at last layer $\ell$, it may start sliding around $C$ and stops after
getting its final position. During the motion of $\cal R$, it remains inside the
circle of radius $d$. Therefore, $\cal R$ remains at minimum distance from
$C$, among all robots in set $R$. After it reaches its final position, it is
removed from set $R$ and added to set $\cal P$. A new nearest robot in set $R$ is
selected.
\qed

\begin{lemma}
No obstacle appears in the path of a mobile robot to its destination.
\end{lemma}

\begin{figure}[!h]
\centering
\includegraphics[height=50mm, width=50mm]{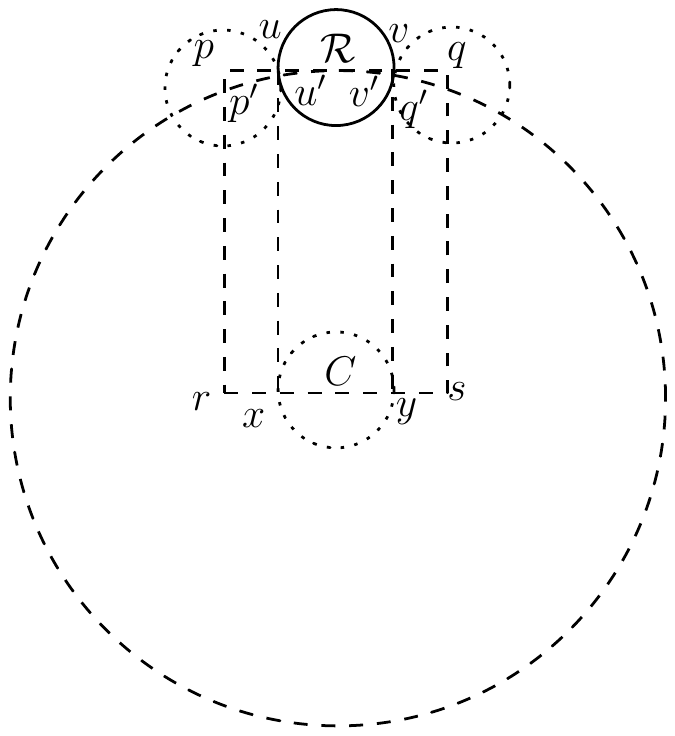}
\caption{No obstacle appears in the path of a mobile robot to its destination}
\label{obstacle}
\end{figure}

\emph{Proof:}
Let $\cal R$ be a robot which is selected for moving towards $C$. According to
$GatheringPattern(\cal R)$, $\cal R$ is nearest to $C$. Let $uxyv$ in Figure. \ref{obstacle} be the
rectangle in which $\cal R$ resides during its movement to $C$. $\overline{uv}$ is
$2$ units. Let rectangles $prxus$ and $vysq$ be of unit width and length equal
to
the length of rectangle $uxyv$. It is sufficient to
show that, if center of no other robot appears inside the quadrilateral $prsq$,
then the path of $\cal R$ towards $C$ will be free of obstacles. If the center
of any other robot appears inside the region $p'rsq'v'u'$, then that robot will
be nearest to $C$. However, $\cal R$ has been selected for moving. Therefore,
$\cal R$ is nearest to $C$ and no robots appear inside the region $p'rsq'v'u'$. 

If any robot appears inside the regions $pp'u'u$ or $vv'q'q$ then either the
robots touch $\cal R$ or is farther from $C$ than $\cal R$. As we consider the
robots in set $R$ are separated, no other robots will touch $\cal R$. According
to $GatheringPattern(\cal R)$, $\cal R$ is the only robot to be considered for moving.
Therefore, any robot inside the regions $pp'u'u$ and/or $vv'q'q$ is not
considered in the path of $\cal R$ to its destination. Hence, no obstacle
appears in the path of a mobile robot to its destination.
\qed

\section{Conclusion}
\label{con}
In this paper, we have proposed a deterministic distributed algorithm for
gathering $n$ $(n \ge 5)$ autonomous, oblivious, homogeneous, asynchronous,
transparent fat robots. The robots do not have explicit communication
capability or common coordinate system. We assume that initially all robots are
stationary and separated. Given a set of such robots, we find a destination
which
remains invariant during the run time of the algorithm. One robot
is allowed to move at any point of time. The robot which moves, is selected such a way that, 
it will be the only eligible robot until it reaches to its destination. 
Therefore, a leader election technique is required to elect a robot for movement from the set 
of homogeneous robots. 
We also propose the leader election algorithm in order to select an robot for movement.

This paper characterizes all the cases where leader election is not possible. To do so, we make an ordering with a set of arbitrarily positioned robots. 
If ordering is possible with a set of robots, then leader election is also possible and hence the robots can gather deterministically. 
The paper characterizes the configurations formed by robots where ordering and hence, gathering is not possible. This configurations are defined as 
symmetric configurations. The gathering algorithm ensures that
multiple mobile robots will gather for all asymmetric configurations.


\end{document}